\documentclass[aip,jmp,preprint,onecolumn,floatfix,superscriptaddress,nofootinbib,11pt]{revtex4}
\usepackage[T1]{fontenc}
\usepackage{lmodern}
\usepackage{amsmath,amssymb,amsthm}
\usepackage{graphicx}
\usepackage[colorlinks=true]{hyperref}
\AtBeginDocument{% optionally set colors to your liking
	\hypersetup{
		urlcolor=cyan,
		citecolor=magenta,
		linkcolor=blue,
		anchorcolor=green,
	}%
}
\usepackage{booktabs,array}
\usepackage{float}

\theoremstyle{plain}
\newtheorem*{maintheorem}{Main Theorem}
\newtheorem{theoremA}{Theorem}

\newtheorem{theoremB}{Theorem}

\newtheorem{propositionC}{Proposition}

\newtheorem{corollaryD}{Corollary}

\newtheorem{propositionE}{Proposition}

\newtheorem{lemma}{Lemma}
\theoremstyle{remark}
\newtheorem*{remark}{Remark}
\numberwithin{equation}{section}
\newcommand{\dd}{\mathrm{d}}
\newcommand{\E}{\mathbb{E}}
\newcommand{\eps}{\varepsilon}
\newcommand{\Law}{\operatorname{Law}}

\begin{document}

\title{A feedback-driven central limit theorem with $q$-Gaussian limits}

\author{G.~Baris Bagci}
\affiliation{Department of Physics, Mersin University, Mersin 33343, Turkey}
\author{Thomas Oikonomou}
\email{thomas.oikonomou1@gmail.com}
\affiliation{Institute of Nanoscience and Nanotechnology, National Center for Scientific Research ``Demokritos'', 15341 Agia Paraskevi, Athens, Greece}

\begin{abstract}
We establish a central limit theorem, the feedback-driven central limit theorem, for sums whose summands are independent noise variables multiplied by an amplitude that depends on the preceding normalized sum $y$. The summands are martingale differences whose conditional variance, $c+b\,y^{2}$, depends quadratically on $y$, with feedback strength $b$; unlike in the martingale central limit theorem, this conditional variance does not settle to a constant. The noise has finite variance; for positive feedback we assume, as in Lyapunov's theorem, a finite third moment, and for negative feedback bounded noise, so that the conditional variance stays nonnegative. For positive feedback, the normalized sum converges to a Student law, equivalently a $q$-Gaussian with $1<q<2$, independently of the noise distribution. For negative feedback, it converges to a compactly supported density, which is a $q$-Gaussian with $q<1$ when $-\tfrac12<b<0$. The Gaussian law, $q=1$, is recovered at zero feedback, while the Cauchy law, $q=2$, is obtained only as a limiting case, in a triangular array in which the feedback grows slowly with the number of terms. Thus every $q<2$ is obtained for a fixed feedback strength, and $q=2$ as a limit. We further show that the balance identity behind the theorem singles out quadratic feedback: a $q$-Gaussian with $q<2$ satisfies it only for a quadratic conditional variance, and no $q$-Gaussian with $2\le q<3$ satisfies it for any conditional variance. The proofs use the method of moments for negative feedback and a Wasserstein coupling argument for positive feedback, including cases in which the limiting variance is infinite. This theorem paves the way to distinguish power laws originating from L\'evy stable laws, whose tails are inherited from heavy-tailed summands, from those originating from $q$-Gaussians, whose tails are generated by feedback while the typical size of the sum, measured by its quantiles, still grows like the square root of the number of terms.
\end{abstract}

\maketitle

\section{Introduction}\label{sec:intro}

A central limit theorem states that a sum of many small random contributions, centered and divided by a suitable scale, has a limit distribution that does not depend on the details of the individual contributions. This universality explains why the Gaussian law appears whenever a quantity is the accumulated result of many independent effects, from errors of measurement to the displacement of a Brownian particle.

The first result of this kind is de Moivre's normal approximation to the binomial distribution of 1733, which Laplace extended in 1812 to sums of more general independent variables~\cite{Fischer}. Proofs for general independent summands followed in the late nineteenth and early twentieth centuries. Chebyshev and Markov introduced the method of moments~\cite{Fischer}; Lyapunov proved the theorem with characteristic functions under a condition on moments of order larger than two~\cite{Lyapunov}; and Lindeberg found the condition, now named after him, under which the Gaussian limit holds~\cite{Lindeberg}. Feller and L\'evy showed that this condition is also necessary when each summand is negligible compared with the sum~\cite{Feller1936,Levy1937}. When the summands have infinite variance, the Gaussian limit is replaced by the stable laws introduced by L\'evy~\cite{Levy1937}, and the generalized central limit theorem of Gnedenko and Kolmogorov characterizes the summands attracted to each stable law through their tails~\cite{GK}. For dependent summands, the martingale central limit theorems~\cite{Brown1971,McLeish1974} give Gaussian limits when the summands are martingale differences whose normalized conditional variances settle down to a constant. In all these theorems the limit is the Gaussian or a stable law, and it depends on the summands only through a few parameters~\cite{Feller}.

In nonextensive statistical mechanics a central role is played by the $q$-Gaussians~\cite{Tsallis1988,Tsallis}. They form a one-parameter family that contains the Gaussian, compactly supported laws and the Student laws, they maximize the Tsallis entropy under suitable constraints, and distributions of this form have been reported in a wide range of systems, among them cold atoms in optical lattices, turbulent flows and the solar wind~\cite{Tsallis}. This raised the question whether the $q$-Gaussians are attractors of sums in the way the Gaussian is. A $q$-generalized central limit theorem was proposed by Umarov, Tsallis and Steinberg~\cite{UTS2008}; it is formulated through a $q$-deformed Fourier transform and a $q$-deformed notion of independence. The $q$-Gaussians considered here should not be confused with the $q$-Gaussian processes of noncommutative probability and their classical versions~\cite{Bryc2001}, a different family that only shares the name.

In this paper we prove a central limit theorem of a different kind, which we call the feedback-driven central limit theorem. It relies on no $q$-deformation of any sort, whether of the Fourier transform, of independence or of the algebra of sums and products. The $q$-Gaussians with $q>1$ are known to arise as limits of sums of exchangeable variables, as Gaussian mixtures whose mixing law is chosen accordingly~\cite{HJU2010}. Here, by contrast, the summands are built from independent noise, and the $q$-Gaussian, above and below $q=1$, is selected by the feedback alone, whatever the law of the noise. To our knowledge, this is the first limit theorem in which $q$-Gaussians arise in this way. The summands are explicitly constructed and uncorrelated: each is a fresh noise variable, independent of the past, multiplied by an amplitude that depends on the current sum divided by the square root of the number of terms, so that the conditional variance of the next summand depends quadratically on this normalized sum. With the normalization of the ordinary central limit theorem, by the square root of the number of terms, the sum converges in distribution to a limit law that is fixed by the strength of the feedback alone; under mild conditions on the noise, the limit does not depend on its law. Positive feedback gives the Student laws, that is, the $q$-Gaussians between the Gaussian and the Cauchy law, including those with infinite variance; negative feedback with bounded noise gives compactly supported laws, which are $q$-Gaussians when the feedback is weak; no feedback gives the Gaussian of the ordinary central limit theorem; and a feedback that grows slowly with the number of terms gives the Cauchy law. The proofs use only standard probability: the method of moments for negative feedback, and a coupling argument in the Wasserstein distance, with an explicit rate of convergence, for positive feedback. Both rest on a single identity, which expresses the balance at every point of the limit density between the contraction due to the normalization and the spreading due to the next summand.

The paper is organized as follows. Section~\ref{sec:1} states the limit theorem, fixes the notation, illustrates the theorem by simulation, reduces its proof to four results and proves the identity satisfied by the limiting density, whose Fourier form is given in Sec.~\ref{sec:2_4}. Sections~\ref{sec:3}--\ref{sec:5} prove these four results, including the Cauchy triangular-array limit, and Section~\ref{sec:concl} contains concluding remarks. The proof of the one-step defect estimate, Proposition~\ref{prop:E} and the details of the numerical checks are given in Appendices~\ref{app:defect}--\ref{app:num}.

\section{Statement}\label{sec:1}

\subsection{The theorem}\label{sec:1_1}

\begin{maintheorem}[feedback-driven central limit theorem]\label{MainTheorem}

Let $\xi_1,\xi_2,\dots$ be independent, identically distributed real random variables with $\E[\xi_1]=0$ and $\E[\xi_1^{2}]=1$, and let $c>0$ and $b\in\mathbb R$. Define random variables $X_1,X_2,\dots$ and their partial sums $S_n$ by
\begin{equation}
\begin{gathered} X_1=\sqrt c\;\xi_1,\qquad X_{n+1}=\sqrt{\,c+b\,\frac{S_n^{2}}{n}\,}\;\xi_{n+1}\quad(n\ge1),\\[4pt] S_n=X_1+X_2+\dots+X_n . \end{gathered}
\label{eq:1.1}
\end{equation}
Assume in addition that $\E[|\xi_1|^{3}]<\infty$ if $b>0$, a moment condition of Lyapunov type, and that $|\xi_1|\le K$ almost surely for a constant $K$ with $|b|\,K^{2}\le1$ if $b<0$, so that the conditional variance $c+b\,S_n^{2}/n$ stays nonnegative. Then, as $n\to\infty$, $S_n/\sqrt n$ converges in distribution to the law with density
\begin{equation}
g_b(y)=\frac{1}{Z_b}\,\big(c+b\,y^{2}\big)^{-1-\frac{1}{2b}}\ \ \text{where } c+b\,y^{2}>0,\qquad g_b(y)=0\ \ \text{elsewhere},
\label{eq:1.2}
\end{equation}
with $Z_b$ the constant that makes the integral equal to $1$, if $b\ne0$, and to the Gaussian law with mean $0$ and variance $c$ if $b=0$. For $b>-\tfrac12$ the limit is the $q$-Gaussian
\begin{equation}
g_b(y)\ \propto\ \Big[1-(1-q)\,B\,y^{2}\Big]_{+}^{\frac{1}{1-q}},\qquad q=\frac{1+4b}{1+2b},\qquad B=\frac{1+2b}{2c},
\label{eq:1.3}
\end{equation}
where $q=1$ ($b=0$) is read as the Gaussian $e^{-By^{2}}$. The limit does not depend on the law of $\xi_1$.

Finally, the Cauchy law $q=2$ is the limit when $b$ grows with the number of summands: if $\E[|\xi_1|^{3}]<\infty$ and the summands \eqref{eq:1.1} are formed with $b=b_N$ and with $c\,(1+b_N)$ in place of $c$, where $b_N\to\infty$ and $b_N^{3/2}\le\theta\ln N$ for all large $N$, for some $0<\theta<\big[8\sqrt2\,(\zeta(\tfrac32)-1)\,\E[|\xi_1|^{3}]\big]^{-1}$, where $\zeta$ is the Riemann zeta function, then $S_N/\sqrt N$ converges in distribution to the Cauchy law with density $\sqrt c/\big(\pi(c+y^{2})\big)$.

\end{maintheorem}

The notation used throughout the manuscript is as follows. We call the variables $\xi_n$ the noise, write $\Law(Y)$ for the distribution of a random variable $Y$, and call $b$ the feedback strength; the feedback is positive for $b>0$ and negative for $b<0$. The conditional variance of the next summand, written as a function of the current normalized sum $y=S_n/\sqrt n$, is denoted by $D(y)=c+b\,y^{2}$, and its square root, wherever $D(y)\ge0$, by $\sigma(y)=\sqrt{D(y)}$. The limit density is a power of this function: with $\kappa=1+1/(2b)$, Eq.~\eqref{eq:1.2} reads $g_b=Z_b^{-1}D^{-\kappa}$. It is positive on the interval $I_b$ where $D>0$, which is the whole line, $I_b=\mathbb R$, for $b>0$, and the bounded interval $I_b=(-L,L)$ with half-width $L=\sqrt{c/|b|}$ for $b<0$. The normalizing constant $Z_b=\int_{I_b}D^{-\kappa}\,\dd y$ is finite: for $b>0$ the integrand is bounded by $c^{-\kappa}$ and decays like $|y|^{-2\kappa}=|y|^{-2-1/b}$, and for $b<0$ it equals $|b|^{-\kappa}\big[(L-y)(L+y)\big]^{1/(2|b|)-1}$, whose exponent is larger than $-1$.

The $q$-Gaussian form \eqref{eq:1.3} is the same density written differently. Since $1-q=-2b/(1+2b)$, one has $(1-q)B=-b/c$ and $1/(1-q)=-\kappa$, so that $\big[1-(1-q)By^{2}\big]^{1/(1-q)}=c^{\kappa}D(y)^{-\kappa}$; conversely, $b=(q-1)/\big(2(2-q)\big)$. Two further standard forms are used in the proofs and in the simulations. For $b>0$, $g_b$ is the law of $\sqrt{c/(1+b)}\;T_\nu$, where $T_\nu$ has Student's law with $\nu=1+1/b$ degrees of freedom; for $b<0$, it is the law of $L(2V-1)$, where $V$ has the Beta law with both parameters equal to $1/(2|b|)$. The resulting regimes are listed in Table~\ref{tab:regimes}.

The proofs work with the normalized sums $Y_n=S_n/\sqrt n$. Dividing $S_{n+1}=S_n+X_{n+1}$ by $\sqrt{n+1}$ turns \eqref{eq:1.1} into the recursion
\begin{equation}
Y_{n+1}=\rho_n\,Y_n+\eps_n\,\sigma(Y_n)\,\xi_{n+1},\qquad \rho_n=\sqrt{\frac{n}{n+1}},\quad \eps_n=\frac{1}{\sqrt{n+1}},
\label{eq:1.4}
\end{equation}
with $\rho_n^{2}+\eps_n^{2}=1$: each step shrinks the current value by the factor $\rho_n$ and adds fresh noise whose standard deviation, $\eps_n\sigma(Y_n)$, depends on that value.
\begin{table}[t]
\caption{Limiting laws for the feedback strength $b$.}
\label{tab:regimes}
\centering\small
\begin{tabular}{lll}
\toprule
$b$ & $q$ & Limit law \\
\midrule
$b\to\infty$ (triangular array) & $2$ & Cauchy \\
$b>0$ & $1<q<2$ & Student; finite variance iff $b<1$ \\
$b=0$ & $1$ & Gaussian \\
$-\tfrac12<b<0$ & $q<1$ & Compactly supported $q$-Gaussian \\
$b=-\tfrac12$ & $q\to-\infty$ & Uniform on $(-L,L)$ \\
$-1\le b<-\tfrac12$ & --- & U-shaped; arcsine at $b=-1$ \\
\bottomrule
\end{tabular}
\end{table}

\begin{figure}[t]
\centering
\includegraphics[width=\textwidth]{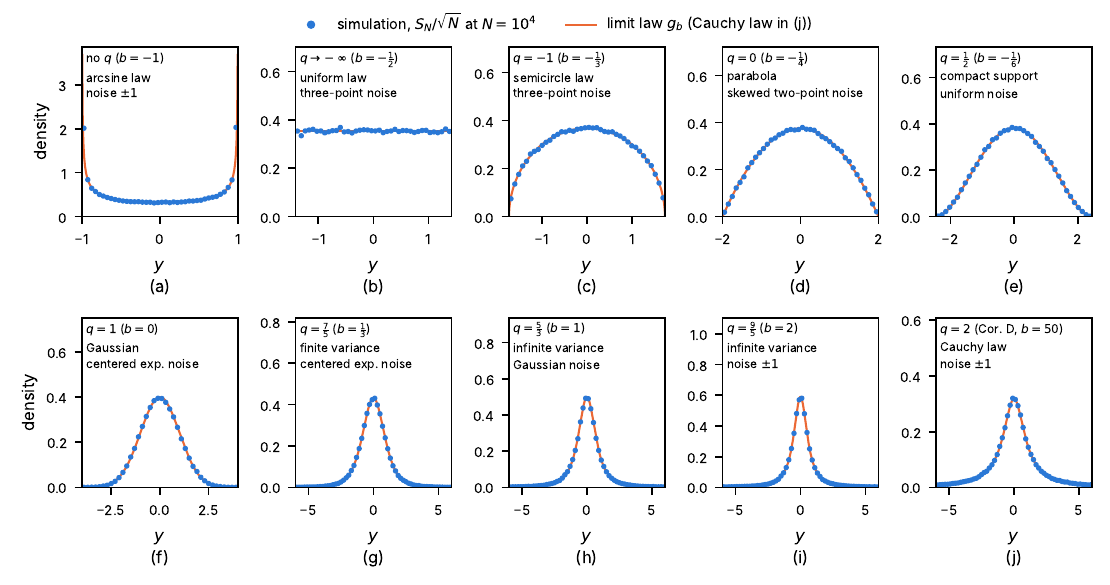}
\caption{Histogram of $S_N/\sqrt N$ at $N=10^{4}$ (points) and the limit law $g_b$ (line; the Cauchy law in panel (j)), with $c=1$ ($c=1+b$ in panel (j)). Each panel gives the index $q=(1+4b)/(1+2b)$, the feedback strength $b$ and the noise.}
\label{fig:1}
\end{figure}

Fig.~\ref{fig:1} illustrates the main theorem by simulation. The recursion \eqref{eq:1.1} was simulated with $c=1$, except in panel (j), using $200{,}000$ independent paths up to $N=10^{4}$. The figure compares the histogram of $S_N/\sqrt N$ at $N=10^{4}$ with the limit law $g_b$ in ten cases, one for each interval and each boundary value of the index $q$: the range $-1\le b<-1/2$, which has no $q$; $q\to-\infty$; $q<0$; $q=0$; $0<q<1$; $q=1$; $1<q<5/3$, with finite variance; the boundary $q=5/3$ and the window $5/3<q<2$, with infinite variance; and the Cauchy law $q=2$ of the triangular array, represented in panel (j) by a run with $b=50$ and $c=1+b$, for which $g_b$ is within $0.002$ of the Cauchy law in Kolmogorov distance. The agreement is at the resolution of the histograms in all panels. The simulation setup, the Kolmogorov distances to $g_b$, the tails and the approach to $q=2$ are reported in Appendix~\ref{app:num}.

\subsection{Proof of the Theorem}\label{sec:1_2}

The main theorem in \ref{MainTheorem} combines the following results. Theorem~\ref{theor:A} is proved in Sec.~\ref{sec:3}, Theorem~\ref{theor:B} in Sec.~\ref{sec:4}, and Proposition~\ref{prop:C} and Corollary~\ref{Corlry:D} in Sec.~\ref{sec:5}. Theorem~\ref{theor:B} is the case $b>0$ and adds a rate of convergence. Theorem~\ref{theor:A} is the case $b<0$ and holds for general summands with the same conditional mean and variance and bounded conditional moments. Proposition~\ref{prop:C} shows that, for fixed $b$, no index $q\ge2$ occurs, and Corollary~\ref{Corlry:D} is the case $q=2$. The case $b=0$ is the classical theorem and is not reproved.

\begin{theoremA}[bounded support, $b<0$]\label{theor:A}

Let $b<0$ and $\beta=-b$. Let $(X_n)_{n\ge1}$ be real random variables adapted to a filtration $(\mathcal F_n)_{n\ge0}$, let $S_0=0$ and $S_n=X_1+\dots+X_n$, and suppose that almost surely:

\begin{itemize}
\item (A1) $\E[X_{n+1}\mid\mathcal F_n]=0$ for all $n\ge0$;
\item (A2) $\E[X_{n+1}^{2}\mid\mathcal F_n]=c-\beta\,S_n^{2}/n$ for all $n\ge1$;
\item (A3) for every integer $j\ge3$ there is a constant $C_j$ with $\E[\,|X_{n+1}|^{j}\mid\mathcal F_n]\le C_j$ for all $n\ge0$.
\end{itemize}
Then $S_n/\sqrt n$ converges in distribution to $g_b$.

The summands \eqref{eq:1.1} satisfy (A1)--(A3) whenever $|\xi_1|\le K$ almost surely for a constant $K$ with $\beta K^{2}\le1$.

\end{theoremA}

\begin{theoremB}[heavy tails, $b>0$]\label{theor:B}

Let $b>0$, let $S_n=X_1+\dots+X_n$ be the sums of the summands \eqref{eq:1.1}, and suppose $M_3=\E[|\xi_1|^{3}]<\infty$. Let $W_1$ be the Wasserstein distance of order 1 (Sec.~\ref{sec:4_1}). Then for every $N\ge3$
\begin{equation}
W_1\Big(\Law\big(S_N/\sqrt N\big),\,g_b\Big)\ \le\ \Lambda\left[\sqrt{\frac{3}{N}}\;W_1\Big(\Law\big(S_3/\sqrt 3\big),\,g_b\Big)+C_L\,\frac{\ln (N/3)}{\sqrt N}\right],
\label{eq:1.5}
\end{equation}
with constants $\Lambda$ and $C_L$ that depend only on $b$, $c$ and $M_3$; they are given in \eqref{eq:4.4} and \eqref{eq:4.7}. In particular $S_N/\sqrt N$ converges in distribution to $g_b$. No finite variance of the limit is needed: the statement holds for every $b>0$, that is, for every $1<q<2$.

\end{theoremB}

\begin{propositionC}[the range of $q$]\label{prop:C}

The map $b\mapsto q=(1+4b)/(1+2b)$ is an increasing bijection from $(0,\infty)$ onto $(1,2)$, and from $(-\tfrac12,0)$ onto $(-\infty,1)$. For no value of $b$ is the limit $g_b$ a $q$-Gaussian on the whole line with $2\le q<3$. In particular the Cauchy law ($q=2$) is not the limit of $S_n/\sqrt n$ for any fixed $b$.

\end{propositionC}

\begin{corollaryD}[the Cauchy law, $q=2$]\label{Corlry:D}

Let $c>0$, let $\xi_1,\xi_2,\dots$ be as in Sec.~\ref{sec:1_1} with $M_3=\E[|\xi_1|^{3}]<\infty$, and let

\begin{equation}
\theta_*=\frac{1}{8\sqrt2\,\big(\zeta(\tfrac32)-1\big)\,M_3}\approx\frac{0.0548}{M_3},
\label{eq:1.6}
\end{equation}

where $\zeta$ is the Riemann zeta function. For each $N\ge3$ choose $b_N>0$, put $c_N=c\,(1+b_N)$, and form the summands \eqref{eq:1.1} with $b=b_N$ and $c=c_N$:

\begin{equation*}
X^{(N)}_1=\sqrt{c_N}\,\xi_1,\qquad X^{(N)}_{n+1}=\sqrt{\,c_N+b_N\,\frac{\big(S^{(N)}_n\big)^{2}}{n}\,}\;\xi_{n+1},\qquad S^{(N)}_n=X^{(N)}_1+\dots+X^{(N)}_n .
\end{equation*}

If $b_N\to\infty$ and $b_N^{3/2}\le\theta\ln N$ for all large $N$, with a constant $0<\theta<\theta_*$, then $S^{(N)}_N/\sqrt N$ converges in distribution to the Cauchy law with density $\sqrt c/\big(\pi(c+z^{2})\big)$, the $q$-Gaussian with $q=2$.

\end{corollaryD}

Along such a sequence $q_N=(1+4b_N)/(1+2b_N)\to2$. Corollary~\ref{Corlry:D} is a statement about a triangular array, in which the rule changes with the number of terms; the Poisson limit of binomial laws with success probability $p_N\to0$ is a classical theorem of the same kind. It does not reach $q>2$.

The two proofs use different methods. Theorem~\ref{theor:A} is proved by moments, which works because the limit has bounded support. Theorem~\ref{theor:B} is proved by a coupling argument in the distance $W_1$, which needs only a first moment and therefore also covers limits with infinite variance. Corollary~\ref{Corlry:D} follows from Theorem~\ref{theor:B} because the constants in \eqref{eq:1.5} are explicit.

\subsection{An identity for the limiting density}\label{sec:2}

\begin{lemma}\label{lem:1}

For every $b\ne0$, on $I_b$,

\begin{equation}
\big(D\,g_b\big)'(y)=-\,y\,g_b(y),
\label{eq:2.1}
\end{equation}

and $D(y)\,g_b(y)\to0$ as $y$ tends to either end of $I_b$.

\end{lemma}

\begin{proof} By \eqref{eq:1.2}, $D\,g_b=Z_b^{-1}D^{\,1-\kappa}$ and $1-\kappa=-\dfrac{1}{2b}$. Since $D'(y)=2by$,

\begin{equation*}
\big(D\,g_b\big)'=\frac{1}{Z_b}\,(1-\kappa)\,D^{-\kappa}\cdot 2by=-\frac{1}{2b}\cdot 2by\cdot g_b=-\,y\,g_b .
\end{equation*}

If $b>0$, then $D^{1-\kappa}=D^{-1/(2b)}\to0$ as $|y|\to\infty$. If $b<0$, then $D^{1-\kappa}=D^{1/(2|b|)}\to0$ as $|y|\to L$, because $D(\pm L)=0$. 
\end{proof}

Identity \eqref{eq:2.1} is the reason $g_b$ is the limit. In the normalized form \eqref{eq:1.4}, a step pulls $Y_n$ toward $0$ by the factor $\rho_n\approx1-\tfrac{1}{2(n+1)}$ and adds noise of variance $D(Y_n)/(n+1)$. A density $g$ is unchanged to first order in $1/n$ by these two effects exactly when $\tfrac12(Dg)''+\tfrac12(yg)'=0$, and \eqref{eq:2.1} gives this.

This first-order condition is the stationary Fokker--Planck equation of the diffusion $\dd U_t=-\tfrac12\,U_t\,\dd t+\sigma(U_t)\,\dd\mathcal W_t$, where $\mathcal W_t$ is a standard Brownian motion, and \eqref{eq:2.1} states that the probability current of this diffusion vanishes for the density $g_b$. Thus $g_b$ is a stationary density of the diffusion, and \eqref{eq:1.4} may be read as a discretization of it with the decreasing step $1/(n+1)$: since $\rho_n=1-1/\big(2(n+1)\big)+O(n^{-2})$,
\begin{equation*}
Y_{n+1}=Y_n-\frac{Y_n}{2(n+1)}+\frac{\sigma(Y_n)\,\xi_{n+1}}{\sqrt{n+1}}+O(n^{-2})\,Y_n .
\end{equation*}
Recursions of this kind, without the $O(n^{-2})$ term, have been shown to converge in distribution to the invariant law of the corresponding diffusion when the diffusion coefficient is Lipschitz and nonvanishing, a quadratic Lyapunov condition holds, and the noise is a martingale difference sequence with unit conditional variance and bounded conditional moments of some order larger than two~\cite{BHW}. Since the Lyapunov expression $2y\cdot(-\tfrac12y)+D(y)$ equals $c-(1-b)\,y^{2}$, these conditions are met for $0<b<1$ under the assumptions of Theorem~\ref{theor:B}, so that this result independently supports Theorem~\ref{theor:B} in the finite-variance range $1<q<5/3$.

\subsection{Fourier form of the identity}\label{sec:2_4}

The Gaussian limit of the ordinary central limit theorem is usually derived with characteristic functions, and the identity \eqref{eq:2.1} has a counterpart of this kind. Write $\varphi_\xi(u)=\E[e^{iu\xi_1}]$ for the characteristic function of the noise. For $b=0$ the summands $X_n=\sqrt c\,\xi_n$ are independent, the characteristic function of their sum is a product, and
\begin{equation}
\E\big[e^{itS_n/\sqrt n}\big]=\varphi_\xi\Big(\frac{\sqrt c\,t}{\sqrt n}\Big)^{n}=\Big(1-\frac{c\,t^{2}}{2n}+o\Big(\frac1n\Big)\Big)^{n}\ \longrightarrow\ e^{-c\,t^{2}/2}.
\label{eq:F1}
\end{equation}
For $b\ne0$ the summands are uncorrelated but not independent, since the amplitude of $X_{n+1}$ depends on $S_n$, and the characteristic function of the sum does not factorize. The computation can nevertheless be carried out one step at a time. Let $\varphi_n(t)=\E[e^{itY_n}]$. In \eqref{eq:1.4} the noise $\xi_{n+1}$ is independent of $Y_n$, so conditioning on $Y_n$ gives the exact relation
\begin{equation}
\varphi_{n+1}(t)=\E\Big[e^{it\rho_nY_n}\,\varphi_\xi\big(t\,\eps_n\,\sigma(Y_n)\big)\Big].
\label{eq:F2}
\end{equation}
For $b=0$ the amplitude $\sigma=\sqrt c$ is constant, the expectation factorizes, and \eqref{eq:F2} is the step that, iterated, gives \eqref{eq:F1}. For $b\ne0$ we insert $\varphi_\xi(u)=1-\tfrac12u^{2}+O(|u|^{3})$ and $\rho_n=1-\tfrac12\eps_n^{2}+O(n^{-2})$, with $\eps_n^{2}=1/(n+1)$. The second-order term contains $\sigma(Y_n)^{2}=c+b\,Y_n^{2}$, and multiplication by $y^{2}$ becomes $-\dd^{2}/\dd t^{2}$ under the Fourier transform, $\E[Y_n^{2}e^{itY_n}]=-\varphi_n''(t)$; this derivative exists for every $n$, because each $Y_n$ has finite variance. Hence, to first order in $1/n$,
\begin{equation}
\varphi_{n+1}(t)=\varphi_n(t)-\frac{1}{2(n+1)}\Big[t\,\varphi_n'(t)+t^{2}\big(c\,\varphi_n(t)-b\,\varphi_n''(t)\big)\Big]+\dots
\label{eq:F3}
\end{equation}
The first term in the bracket comes from the contraction by $\rho_n$, the second from the spreading by the next summand. A characteristic function is unchanged to this order exactly when the bracket vanishes, that is, for $t\ne0$,
\begin{equation}
b\,t\,\varphi''(t)-\varphi'(t)-c\,t\,\varphi(t)=0,\qquad \varphi(0)=1 .
\label{eq:F4}
\end{equation}
This is the Fourier transform of \eqref{eq:2.1}: the transform of $(Dg)'$ is $-it\,(c\,\varphi-b\,\varphi'')$, and that of $-y\,g$ is $i\,\varphi'$. For $b\ge1$ the function $D\,g_b$, which decays like $|y|^{-1/b}$, is not integrable, and the transform is then taken in the sense of distributions, or equivalently \eqref{eq:F4} is read for $t\ne0$ only.

Equation \eqref{eq:F4} is a Bessel equation, and its solutions with $\varphi(0)=1$ and $|\varphi|\le1$ are the characteristic functions of the limit laws $g_b$. The equation is invariant under $t\mapsto-t$ and is solved separately for $t>0$ and $t<0$; both half-lines give the same function of $|t|$, so the solution is real and even, as it must be for the symmetric law $g_b$, even when the noise is skewed and the $\varphi_n$ are complex.
\begin{itemize}
\item For $b=0$ it reduces to $\varphi'=-c\,t\,\varphi$, with solution $e^{-ct^{2}/2}$, as in \eqref{eq:F1}.
\item For $b>0$, with $\nu=1+1/b$ and $k=\sqrt{c/b}$,
\begin{equation*}
\varphi(t)=\frac{2^{1-\nu/2}}{\Gamma(\nu/2)}\,(k|t|)^{\nu/2}\,K_{\nu/2}(k|t|),
\end{equation*}
where $K_{\nu/2}$ is the modified Bessel function of the second kind. This is the characteristic function of $\sqrt{c/(1+b)}\;T_\nu$, Student's law with $\nu$ degrees of freedom and the scale of Sec.~\ref{sec:1_1}, since $\sqrt\nu\,\sqrt{c/(1+b)}=k$. The second solution, $(k|t|)^{\nu/2}I_{\nu/2}(k|t|)$, grows like $e^{k|t|}$ and is excluded by $|\varphi|\le1$.
\item For $-1\le b<0$, with $a=1/(2|b|)$ and $L=\sqrt{c/|b|}$,
\begin{equation*}
\varphi(t)=\Gamma\big(a+\tfrac12\big)\Big(\frac{2}{L\,t}\Big)^{a-\frac12}J_{a-\frac12}(L\,t),
\end{equation*}
where $J_{a-1/2}$ is the Bessel function of the first kind. This is the characteristic function of $L(2V-1)$ with $V$ distributed according to the Beta law with both parameters equal to $a$, as in Sec.~\ref{sec:1_1}. The uniform law, $b=-\tfrac12$, gives $\sin(Lt)/(Lt)$, and the arcsine law, $b=-1$, gives $J_0(Lt)$. Since $a-\tfrac12\ge0$ for $b\ge-1$, the second solution, built from the Bessel function $Y_{a-1/2}$, is singular at $t=0$ and is excluded by $\varphi(0)=1$. For $b<-1$, which Theorem~\ref{theor:A} allows for general summands, the order $a-\tfrac12$ is negative and the second solution, $t^{\frac12-a}J_{\frac12-a}(L\,t)$, is bounded and vanishes at $t=0$, so that $\varphi(0)=1$ and $|\varphi|\le1$ no longer exclude adding a multiple of it. Since it behaves like $t^{1-2a}$ with $0<1-2a<1$, it has an infinite derivative at $t=0$, and it is excluded by requiring, in addition, a finite first moment, that is, a finite $\varphi'(0)$.
\end{itemize}
Thus the ordinary Fourier transform turns the balance identity into a single second-order equation, whose bounded solutions are the Gaussian, Student and Beta characteristic functions, in the same way as it turns the classical computation \eqref{eq:F1} into the Gaussian characteristic function.

The two derivations are compared in Table~\ref{tab:fourier}. The classical one transforms the whole sum and uses the independence of all summands, which turns the characteristic function of the sum into a product. The present one transforms a single step and uses only the independence of the next noise variable from the past. The exact step \eqref{eq:F2} does not close on $\varphi_n$, even for quadratic feedback, because the noise is evaluated at $t\,\eps_n\,\sigma(Y_n)$ and $\sigma=\sqrt D$ is not a polynomial; already the third-order term contains $\sigma^{3}=D^{3/2}$. The quadratic form of the conditional variance is what makes the first-order step \eqref{eq:F3} close: for a general variance function $D$, the expectation $\E[D(Y_n)\,e^{itY_n}]$ is not expressed through $\varphi_n$ and finitely many of its derivatives. For a polynomial $D$ it is, and the first-order step closes into an equation whose order is the degree of $D$; for $D=c+by^{2}$ this equation is of second order and is solved explicitly by Bessel functions, and by Proposition~\ref{prop:E} only this choice of $D$ balances a $q$-Gaussian. In both derivations the law of the noise enters at first order only through its variance, the second-order term of $\varphi_\xi$, which is why the limit does not depend on it. Its third moment enters at the next order, $n^{-3/2}$, and affects only the speed of convergence, consistent with the slower approach observed for skewed noise in Table~\ref{tab:1}.

The derivation of \eqref{eq:F4} is formal, and this is the main difference from the classical case. In \eqref{eq:F1} the remainder of the expansion of $\varphi_\xi$ is controlled by the moments of the noise alone. In \eqref{eq:F2} it is multiplied by powers of $\sigma(Y_n)$, so that it is controlled only if $Y_n$ has enough moments, uniformly in $n$. The third-order bound needs $\sup_n\E[|Y_n|^{3}]<\infty$; this fails for $b\ge\tfrac12$, since $g_b$ has an infinite third moment and $Y_n$ converges to $g_b$ in distribution, and we expect, but have not shown, that it holds for $0<b<\tfrac12$. The bound $|\varphi_\xi(u)-1+\tfrac12u^{2}|\le\E\big[\min\big(u^{2}\xi_1^{2},\tfrac16|u\,\xi_1|^{3}\big)\big]$ needs only second moments, but makes the remainder of order $o(1/n)$ only if the $Y_n^{2}$ are uniformly integrable, which can hold only for $b<1$ (see Remark after Proposition~\ref{prop:C}), and summing the remainders over $n$ would in addition require a rate. For $b\ge1$ the limit has infinite variance, $\varphi''(0)$ does not exist, and \eqref{eq:F4} holds only for $t\ne0$. Moreover, passing from \eqref{eq:F3} to \eqref{eq:F4} assumes that not only $\varphi_n$ but also $\varphi_n'$ and $\varphi_n''$ converge, whereas convergence in distribution gives only the pointwise convergence of $\varphi_n$; and turning \eqref{eq:F3} into a convergence proof would also require the stability of the recursion around the solution of \eqref{eq:F4}. For these reasons \eqref{eq:F4} is used here to identify the limit, and convergence is proved with moments for $b<0$ and with the coupling of Sec.~\ref{sec:4} for $b>0$, which covers infinite variance.

\begin{table}[H]
\caption{Fourier derivation of the ordinary and of the feedback-driven central limit theorem. The row ``Step equation'' refers to the one-step form \eqref{eq:F3}, which for $b=0$ is equivalent to the product \eqref{eq:F1}.}
\label{tab:fourier}
\centering\small
\begin{tabular}{lll}
\toprule
 & Ordinary CLT ($b=0$) & Feedback-driven CLT ($b\ne0$) \\
\midrule
Summands & independent & uncorrelated; amplitude set by $S_n/\sqrt n$ \\
Fourier transform of & the whole sum, Eq.~\eqref{eq:F1} & one step, Eq.~\eqref{eq:F2} \\
Property used & independence of all summands & independence of $\xi_{n+1}$ from the past \\
Second-order term & factor $1-c\,t^{2}/(2n)$ & $-t^{2}\big(c\,\varphi-b\,\varphi''\big)/\big(2(n+1)\big)$ \\
Limit & computed directly from \eqref{eq:F1} & bounded solution of the step equation \\
Step equation & $\varphi'=-c\,t\,\varphi$ & $b\,t\,\varphi''-\varphi'-c\,t\,\varphi=0$ \\
Solution & $e^{-ct^{2}/2}$ & Bessel $K$ ($b>0$), Bessel $J$ ($b<0$) \\
Limit law & Gaussian & Student ($b>0$), Beta ($b<0$) \\
Noise enters through & its variance & its variance (at first order) \\
Status & proof (L\'evy continuity theorem) & formal; proofs in Secs.~\ref{sec:3}, \ref{sec:4} \\
\bottomrule
\end{tabular}
\end{table}

\section{Proof of Theorem~\ref{theor:A}}\label{sec:3}

Throughout this section $b<0$, $\beta=-b>0$, $L=\sqrt{c/\beta}$, and (A1)--(A3) hold. Write $Y_n=S_n/\sqrt n$ for $n\ge1$. We show that the normalized sum is bounded, that each of its moments converges to the corresponding moment of $g_b$, and that this implies convergence in distribution; at the end of the section we show that the summands \eqref{eq:1.1} satisfy (A1)--(A3).

We first show that the normalized sum is bounded. The left side of (A2) is a conditional expectation of a nonnegative variable, so it is nonnegative almost surely. Hence $c-\beta S_n^{2}/n\ge0$, that is,
\begin{equation}
S_n^{2}\le \frac{c\,n}{\beta}\quad\text{and}\quad |Y_n|\le L\qquad\text{almost surely, for all }n\ge1.
\label{eq:3.1}
\end{equation}
Hence the normalized sum is bounded.

Next we derive a recursion for the moments of the sum. Fix an integer $r\ge2$ and $n\ge1$. By the binomial formula,
\begin{equation*}
S_{n+1}^{\,r}=\sum_{j=0}^{r}\binom rj\,S_n^{\,r-j}\,X_{n+1}^{\,j}.
\end{equation*}
Every term is integrable, because $S_n$ is bounded by \eqref{eq:3.1} and $X_{n+1}$ has finite moments of all orders by (A3). Take the conditional expectation given $\mathcal F_n$. The term $j=0$ is $S_n^{r}$. The term $j=1$ vanishes by (A1). The term $j=2$ equals $\binom r2 S_n^{\,r-2}\big(c-\beta S_n^{2}/n\big)$ by (A2). For the terms $j\ge3$, (A3) and \eqref{eq:3.1} give
\begin{equation*}
\Big|\sum_{j=3}^{r}\binom rj S_n^{\,r-j}\,\E\big[X_{n+1}^{\,j}\mid\mathcal F_n\big]\Big|\ \le\ \sum_{j=3}^{r}\binom rj\,(L\sqrt n)^{\,r-j}C_j\ \le\ \bar C_r\,n^{\frac{r-3}{2}},\qquad \bar C_r=\sum_{j=3}^{r}\binom rj L^{\,r-j}C_j ,
\end{equation*}
where $n^{(r-j)/2}\le n^{(r-3)/2}$ was used (for $r=2$ the sum is empty and $\bar C_2=0$). Taking expectations, with $\mathcal M_r(n)=\E[S_n^{\,r}]$,
\begin{equation}
\mathcal M_r(n+1)=\Big(1-\binom r2\frac{\beta}{n}\Big)\mathcal M_r(n)+\binom r2\,c\;\mathcal M_{r-2}(n)+\mathcal R_r(n),\qquad |\mathcal R_r(n)|\le \bar C_r\,n^{\frac{r-3}{2}} .
\label{eq:3.2}
\end{equation}
Hence the moments of the sum obey the recursion \eqref{eq:3.2}.

We now write the same recursion for the moments of the normalized sum. Let $m_r(n)=\E[Y_n^{\,r}]=\mathcal M_r(n)/n^{r/2}$. By \eqref{eq:3.1}, $|m_r(n)|\le L^{r}$. Divide \eqref{eq:3.2} by $(n+1)^{r/2}$ and write $u_n=\big(\tfrac{n}{n+1}\big)^{r/2}$ and $\beta_r=\binom r2\beta$:
\begin{equation*}
m_r(n+1)=u_n\Big(1-\frac{\beta_r}{n}\Big)m_r(n)+u_n\,\binom r2\frac{c}{n}\,m_{r-2}(n)+\frac{\mathcal R_r(n)}{(n+1)^{r/2}} .
\end{equation*}
For $x\in(0,1]$, Taylor's formula with Lagrange remainder applied to $(1+x)^{-r/2}$ gives
\begin{equation*}
\Big|(1+x)^{-r/2}-1+\frac r2\,x\Big|\le\frac12\cdot\frac r2\Big(\frac r2+1\Big)x^{2}.
\end{equation*}
With $x=1/n$ this reads $u_n=1-\dfrac{r}{2n}+\omega_n$ with $|\omega_n|\le\dfrac{r(r+2)}{8n^{2}}$. Therefore
\begin{equation*}
u_n\Big(1-\frac{\beta_r}{n}\Big)=1-\frac{\lambda_r}{n}+\Big[\frac{r\beta_r}{2n^{2}}+\omega_n\Big(1-\frac{\beta_r}{n}\Big)\Big],\qquad \lambda_r=\frac r2+\binom r2\beta=\frac r2\big(1+(r-1)\beta\big)>0 .
\end{equation*}
Collecting terms,
\begin{equation}
m_r(n+1)=\Big(1-\frac{\lambda_r}{n}\Big)m_r(n)+\frac{\lambda_r}{n}\,z_r(n)+e_r(n),\qquad z_r(n)=\frac{(r-1)\,c}{1+(r-1)\beta}\;m_{r-2}(n),
\label{eq:3.3}
\end{equation}
where $\binom r2c/\lambda_r=(r-1)c/(1+(r-1)\beta)$ was used, and the error is
\begin{equation*}
e_r(n)=\Big[\frac{r\beta_r}{2n^{2}}+\omega_n\Big(1-\frac{\beta_r}{n}\Big)\Big]m_r(n)+(u_n-1)\binom r2\frac cn\,m_{r-2}(n)+\frac{\mathcal R_r(n)}{(n+1)^{r/2}} .
\end{equation*}
Each of the three parts is bounded by a constant times $n^{-3/2}$: the first by $\big[\tfrac{r\beta_r}{2}+\tfrac{r(r+2)}{8}(1+\beta_r)\big]L^{r}n^{-2}$; the second by $\tfrac r2\binom r2c\,L^{r-2}n^{-2}$, because $|u_n-1|\le\tfrac{r}{2n}$; the third by $\bar C_r\,n^{(r-3)/2}/n^{r/2}=\bar C_r\,n^{-3/2}$. Hence
\begin{equation}
\sum_{n\ge1}|e_r(n)|<\infty .
\label{eq:3.4}
\end{equation}
Hence the moments of the normalized sum obey a recursion of the form \eqref{eq:3.3} with summable errors. Recursions of this form converge, as the following lemma shows.

\begin{lemma}\label{lem:2}

Let $\lambda>0$ and let $(x_n)$, $(z_n)$, $(e_n)$ be real sequences with
\begin{equation*}
x_{n+1}=\Big(1-\frac\lambda n\Big)x_n+\frac\lambda n\,z_n+e_n\quad(n\ge1),\qquad z_n\to z,\qquad \sum_n|e_n|<\infty .
\end{equation*}
Then $x_n\to z$.

\end{lemma}

\begin{proof} Put $w_n=x_n-z$. Then $w_{n+1}=\big(1-\tfrac\lambda n\big)w_n+\tfrac\lambda n(z_n-z)+e_n$. Fix $\eta>0$ and choose $n_1>\lambda$ such that $|z_n-z|\le\eta$ for all $n\ge n_1$ and $\sum_{n\ge n_1}|e_n|\le\eta$. For $n_1\le n\le N$ let
\begin{equation*}
\Pi_{n,N}=\prod_{k=n}^{N-1}\Big(1-\frac\lambda k\Big)\in(0,1],\qquad \Pi_{N,N}=1 .
\end{equation*}
Iterating the recursion from $n_1$ to $N$,
\begin{equation*}
w_N=\Pi_{n_1,N}\,w_{n_1}+\sum_{n=n_1}^{N-1}\Pi_{n+1,N}\Big[\frac\lambda n(z_n-z)+e_n\Big].
\end{equation*}
Since $\Pi_{n+1,N}-\Pi_{n,N}=\tfrac\lambda n \Pi_{n+1,N}$, the sum $\sum_{n=n_1}^{N-1}\tfrac\lambda n\Pi_{n+1,N}$ telescopes to $1-\Pi_{n_1,N}\le1$. Therefore
\begin{equation*}
|w_N|\le \Pi_{n_1,N}\,|w_{n_1}|+\eta+\eta .
\end{equation*}
Finally $\Pi_{n_1,N}\le\exp\big(-\lambda\sum_{k=n_1}^{N-1}\tfrac1k\big)\to0$ as $N\to\infty$, because the harmonic series diverges. So $\limsup_N|w_N|\le2\eta$ for every $\eta>0$, that is, $w_N\to0$. 
\end{proof}

We now show by induction on $r$ that $m_r(n)\to m_r$ as $n\to\infty$, where
\begin{equation}
m_0=1,\qquad m_1=0,\qquad m_r=\frac{(r-1)\,c}{1+(r-1)\beta}\;m_{r-2}\quad(r\ge2).
\label{eq:3.5}
\end{equation}
For $r=0$ this is trivial. For $r=1$, (A1) gives $\E[S_n]=0$, so $m_1(n)=0$. Let $r\ge2$ and suppose $m_{r-2}(n)\to m_{r-2}$. In \eqref{eq:3.3}, $z_r(n)\to\tfrac{(r-1)c}{1+(r-1)\beta}m_{r-2}=m_r$, and the errors are summable by \eqref{eq:3.4}. Lemma \ref{lem:2} with $\lambda=\lambda_r$ gives $m_r(n)\to m_r$. Hence all moments of the normalized sum converge.

Next we show that the limits $m_r$ are the moments of the limit law. Let $m_r^{g}=\int_{-L}^{L}y^{r}g_b(y)\,\dd y$. Then $m_0^{g}=1$, and $m_1^{g}=0$ because $g_b$ is even. For $r\ge2$, use \eqref{eq:2.1}, integrate by parts, and use that $Dg_b$ vanishes at $\pm L$ (Lemma \ref{lem:1}):
\begin{equation*}
\begin{aligned} m_r^{g}&=\int_{-L}^{L}y^{\,r-1}\cdot y\,g_b\,\dd y=-\int_{-L}^{L}y^{\,r-1}\big(Dg_b\big)'\dd y\\[4pt] &=(r-1)\int_{-L}^{L}y^{\,r-2}\,(c-\beta y^{2})\,g_b\,\dd y=(r-1)\big(c\,m_{r-2}^{g}-\beta\,m_r^{g}\big). \end{aligned}
\end{equation*}
Solving for $m_r^{g}$ gives the recursion \eqref{eq:3.5}. Hence $m_r^{g}=m_r$ for all $r\ge0$, that is, the moments of the normalized sum converge to those of $g_b$.

Finally, convergence of the moments implies convergence in distribution. By \eqref{eq:3.1} every $Y_n$ takes values in $[-L,L]$, and $g_b$ is carried by the same interval. By the two preceding results, $\E[p(Y_n)]\to\int p\,g_b$ for every polynomial $p$. Let $h$ be continuous on $[-L,L]$ and $\eta>0$. By the Weierstrass approximation theorem there is a polynomial $p$ with $|h-p|\le\eta$ on $[-L,L]$. Then
\begin{equation*}
\Big|\E[h(Y_n)]-\int h\,g_b\Big|\le 2\eta+\Big|\E[p(Y_n)]-\int p\,g_b\Big|\ \longrightarrow\ 2\eta .
\end{equation*}
So $\E[h(Y_n)]\to\int h\,g_b$ for every continuous $h$. Hence $S_n/\sqrt n$ converges in distribution to $g_b$. \qed

It remains to show that the summands \eqref{eq:1.1} satisfy the hypotheses (A1)--(A3). Let $|\xi_1|\le K$ almost surely with $\beta K^{2}\le1$, and let $\mathcal F_n$ be generated by $\xi_1,\dots,\xi_n$. We first show by induction that $S_n^{2}\le cn/\beta$ for all $n\ge1$, so that the square root in \eqref{eq:1.1} is defined. For $n=1$: $S_1^{2}=c\,\xi_1^{2}\le cK^{2}\le c/\beta$. Suppose $S_n^{2}\le cn/\beta$. By \eqref{eq:1.1} and the Cauchy--Schwarz inequality,
\begin{equation*}
\begin{aligned} |S_{n+1}|&\le \sqrt{\frac{\beta S_n^{2}}{n}}\cdot\sqrt{\frac n\beta}+\sqrt{c-\frac{\beta S_n^{2}}{n}}\cdot K\ \le\ \sqrt{\frac{\beta S_n^{2}}{n}+c-\frac{\beta S_n^{2}}{n}}\;\sqrt{\frac n\beta+K^{2}}\\[4pt] &=\sqrt{c\Big(\frac n\beta+K^{2}\Big)}\le\sqrt{\frac{c\,(n+1)}{\beta}} , \end{aligned}
\end{equation*}
where the last step uses $K^{2}\le1/\beta$. This closes the induction. Now $X_{n+1}=\sigma(Y_n)\,\xi_{n+1}$, where $\sigma(Y_n)$ is $\mathcal F_n$-measurable and $\xi_{n+1}$ is independent of $\mathcal F_n$. So $\E[X_{n+1}\mid\mathcal F_n]=\sigma(Y_n)\,\E[\xi_{n+1}]=0$, which is (A1); $\E[X_{n+1}^{2}\mid\mathcal F_n]=D(Y_n)=c-\beta S_n^{2}/n$, which is (A2); and $|X_{n+1}|\le\sqrt c\,K$, which gives (A3) with $C_j=(\sqrt c\,K)^{j}$. For $n=0$ the same holds with $X_1=\sqrt c\,\xi_1$. Hence the summands \eqref{eq:1.1} satisfy (A1)--(A3). \qed

\begin{remark}
Since $\E[\xi_1^{2}]=1$ forces $K\ge1$, the condition $\beta K^{2}\le1$ can hold only for $-1\le b<0$. For $b<-1$ the summands \eqref{eq:1.1} are not defined: \eqref{eq:3.1} with $n=1$ would require $\xi_1^{2}\le1/\beta<1$ almost surely, which contradicts $\E[\xi_1^{2}]=1$. Theorem~\ref{theor:A} still applies to any sequence satisfying (A1)--(A3), but no example with $b<-1$ is constructed here. Such a sequence cannot be of the product form $\sigma(Y_n)\,\xi_{n+1}$ with independent, identically distributed $\xi_n$; it would require a noise law that depends on the current sum, for example asymmetric two-point steps that are shorter in the outward direction.
\end{remark}

\section{Proof of Theorem~\ref{theor:B}}\label{sec:4}

Throughout this section $b>0$, $g=g_b$, $\kappa=1+\tfrac{1}{2b}>1$, and $M_3=\E[|\xi_1|^{3}]<\infty$. By Lyapunov's inequality $M_3^{1/3}\ge(\E[\xi_1^{2}])^{1/2}=1$, so $M_3\ge1$; also $\E[|\xi_1|]\le1$.

\subsection{The distance}\label{sec:4_1}

Let $\mathcal P_1$ be the set of probability measures on $\mathbb R$ with finite first moment. For $\mu,\mu'\in\mathcal P_1$,
\begin{equation*}
W_1(\mu,\mu')=\inf\big\{\E[|U-U'|]\ :\ U\sim\mu,\ U'\sim\mu'\big\},
\end{equation*}
where the infimum is over all pairs of random variables on one probability space with the given laws. A density is identified with the measure it defines. Three standard facts are used~\cite{Villani}:

\begin{itemize}
\item (W1) The infimum is attained by some pair $(U,U')$.
\item (W2) $W_1$ is a metric on $\mathcal P_1$, and
\begin{equation}
W_1(\mu,\mu')=\sup\Big\{\int h\,\dd\mu-\int h\,\dd\mu'\ :\ |h(x)-h(y)|\le|x-y|\ \text{for all }x,y\Big\}.
\label{eq:4.1}
\end{equation}

\item (W3) If $W_1(\mu_N,\mu)\to0$, then $\mu_N\to\mu$ in distribution. Indeed, by \eqref{eq:4.1}, $\big|\int h\,\dd\mu_N-\int h\,\dd\mu\big|\le \mathrm{Lip}(h)\,W_1(\mu_N,\mu)$ for every bounded Lipschitz $h$, where $\mathrm{Lip}(h)$ is its Lipschitz constant, and convergence of these integrals for all bounded Lipschitz $h$ is convergence in distribution (portmanteau theorem~\cite{Billingsley}).
\end{itemize}

\begin{lemma}[smooth test functions suffice]\label{lem:3}

Let $\mathcal H$ be the set of $C^\infty$ functions $h:\mathbb R\to\mathbb R$ with $|h'|\le1$ and $h'$ of compact support. For $\mu,\mu'\in\mathcal P_1$,
\begin{equation*}
W_1(\mu,\mu')=\sup_{h\in\mathcal H}\Big(\int h\,\dd\mu-\int h\,\dd\mu'\Big).
\end{equation*}

\end{lemma}

\begin{proof} Every $h\in\mathcal H$ is 1-Lipschitz, so ``$\ge$'' follows from \eqref{eq:4.1}. For ``$\le$'', let $h$ be 1-Lipschitz, $R>0$ and $\delta\in(0,1)$. Let $h_R(y)=h\big(\max(-R,\min(R,y))\big)$. Then $h_R$ is 1-Lipschitz, constant on $(-\infty,-R]$ and on $[R,\infty)$, and $|h-h_R|(y)\le(|y|-R)_+$. Let $\varphi_\delta\ge0$ be a $C^\infty$ function supported in $[-\delta,\delta]$ with integral 1, and $h_{R,\delta}=h_R*\varphi_\delta$. Then $h_{R,\delta}\in\mathcal H$ (its derivative vanishes outside $[-R-\delta,R+\delta]$) and $|h_{R,\delta}-h_R|\le\delta$. Hence
\begin{equation*}
\int h\,\dd\mu-\int h\,\dd\mu'\ \le\ \int h_{R,\delta}\,\dd\mu-\int h_{R,\delta}\,\dd\mu'+2\delta+\int(|y|-R)_+\,\dd(\mu+\mu')(y).
\end{equation*}
The last integral tends to $0$ as $R\to\infty$, because $\mu$ and $\mu'$ have finite first moments. Letting $\delta\to0$ and $R\to\infty$ and taking the supremum over $h$ gives ``$\le$''. 
\end{proof}

\subsection{One step as a map on laws}\label{sec:4_2}

For $\eps\in(0,1)$, $\rho=\sqrt{1-\eps^{2}}$ and $\mu\in\mathcal P_1$, let $\mu \mathcal K^{\eps}$ be the law of
\begin{equation}
\rho\,U+\eps\,\sigma(U)\,\xi,\qquad U\sim\mu,\quad \xi\sim\xi_1,\quad U\text{ and }\xi\text{ independent}.
\label{eq:4.2}
\end{equation}
Since $\sigma(u)\le\sqrt c+\sqrt b\,|u|$, we have $\mu \mathcal K^{\eps}\in\mathcal P_1$. Write $\mathcal K_n=\mathcal K^{\eps_n}$ with $\eps_n=(n+1)^{-1/2}$, and for $m\le N$ let $\mu \mathcal K_{m,N}$ be the result of applying $\mathcal K_m,\mathcal K_{m+1},\dots,\mathcal K_{N-1}$ to $\mu$ in this order ($\mathcal K_{N,N}$ is the identity). In \eqref{eq:1.4}, $\xi_{n+1}$ is independent of $Y_n$, so $\Law(Y_{n+1})=\Law(Y_n)\mathcal K_n$, and therefore
\begin{equation}
\Law(Y_N)=\Law(Y_m)\,\mathcal K_{m,N}\qquad(1\le m\le N).
\label{eq:4.3}
\end{equation}
Also $\Law(Y_n)\in\mathcal P_1$ for every $n$, by induction from $\E[|Y_1|]=\sqrt c\,\E[|\xi_1|]<\infty$.

\subsection{Elementary bounds}\label{sec:4_3}

\begin{lemma}\label{lem:4}

Let $b>0$. For all real $y$:
\begin{itemize}
\item(a) $\sigma(y)\ge\sqrt c$ and $|y|\le\sigma(y)/\sqrt b$;
\item(b) $\sigma'=\dfrac{b\,y}{\sigma}$ and $|\sigma'|\le\sqrt b$; consequently $|\sigma(y)-\sigma(y')|\le\sqrt b\,|y-y'|$;
\item(c) $\sigma''=\dfrac{b\,c}{\sigma^{3}}>0$ and $|\sigma'''|\le\dfrac{3\,b^{3/2}c}{\sigma^{4}}$;
\item(d) $\big|(y\,g)'\big|\le2\kappa\,g$;
\item(e) $a_1:=\displaystyle\int\sigma\,g\,\dd y<\infty$ and $\displaystyle\int|y|\,g\,\dd y\le a_1/\sqrt b$.
\end{itemize}

\end{lemma}

\begin{proof} (a) $\sigma^{2}=c+by^{2}$ is at least $c$ and at least $by^{2}$. (b) Differentiate $\sigma^{2}=c+by^{2}$ to get $\sigma\sigma'=by$; then $|\sigma'|=b|y|/\sigma\le\sqrt b$ by (a), and the Lipschitz bound follows from the mean value theorem. (c) By the quotient rule and (b), $\sigma''=(b\sigma-by\,\sigma')/\sigma^{2}$. Inserting $\sigma'=by/\sigma$ gives $\sigma''=(b\sigma^{2}-b^{2}y^{2})/\sigma^{3}=bc/\sigma^{3}$, and $\sigma'''=-3bc\,\sigma'/\sigma^{4}$, so $|\sigma'''|\le3b^{3/2}c/\sigma^{4}$ by (b). (d) From \eqref{eq:1.2}, $g'=-2\kappa b\,y\,g/D$, so $(yg)'=g\,(1-2\kappa\,by^{2}/D)$; since $0\le by^{2}/D<1$ and $\kappa>1$, the bracket lies in $(1-2\kappa,1]$ and its absolute value is at most $2\kappa-1\le2\kappa$. (e) $\sigma g\le Z_b^{-1}D^{\frac12-\kappa}$, which is bounded and is at most $Z_b^{-1}b^{\frac12-\kappa}|y|^{-1-1/b}$, hence integrable; the second claim follows from (a). 
\end{proof}

\subsection{Two normalized sums driven by the same noise come together}\label{sec:4_4}

\begin{lemma}[contraction]\label{lem:5}

Let
\begin{equation}
\Lambda=\exp\Big(4\sqrt2\;b^{3/2}M_3\sum_{j\ge2}j^{-3/2}\Big).
\label{eq:4.4}
\end{equation}
Then for all integers $1\le m\le N$ and all $\mu,\mu'\in\mathcal P_1$,
\begin{equation}
W_1\big(\mu \mathcal K_{m,N},\,\mu' \mathcal K_{m,N}\big)\ \le\ \Lambda\,\sqrt{\frac mN}\;W_1(\mu,\mu').
\label{eq:4.5}
\end{equation}

\end{lemma}

\begin{proof} We couple the two laws, show that one step of the recursion multiplies the expected distance of the coupled sums by at most $\rho_n(1+\delta_n)$, and iterate. By (W1) choose $(U,U')$ with $U\sim\mu$, $U'\sim\mu'$ and $\E[|U-U'|]=W_1(\mu,\mu')$, independent of $\xi_{m+1},\xi_{m+2},\dots$. Define two sequences driven by the same noise:
\begin{equation*}
\begin{gathered} Y_m=U,\qquad Y'_m=U',\\[2pt] Y_{n+1}=\rho_nY_n+\eps_n\sigma(Y_n)\xi_{n+1},\qquad Y'_{n+1}=\rho_nY'_n+\eps_n\sigma(Y'_n)\xi_{n+1}\qquad(n\ge m). \end{gathered}
\end{equation*}
Then $Y_N\sim\mu \mathcal K_{m,N}$ and $Y'_N\sim\mu' \mathcal K_{m,N}$, so $W_1(\mu \mathcal K_{m,N},\mu' \mathcal K_{m,N})\le\E[|Y_N-Y'_N|]$.

Next we bound one step. Let $\Delta_n=Y_n-Y'_n$ and let $\vartheta_n=\big(\sigma(Y_n)-\sigma(Y'_n)\big)/\Delta_n$ if $\Delta_n\ne0$, and $\vartheta_n=0$ otherwise. By Lemma \ref{lem:4}(b), $|\vartheta_n|\le\sqrt b$. Subtracting the two recursions,
\begin{equation*}
\Delta_{n+1}=\Delta_n\,\big(\rho_n+\eps_n\vartheta_n\,\xi_{n+1}\big).
\end{equation*}
Here $\Delta_n$ and $\vartheta_n$ are functions of $U,U',\xi_{m+1},\dots,\xi_n$, and $\xi_{n+1}$ is independent of these. For a real number $x$ write $x^-=\max(-x,0)$, so that $|x|=x+2x^-$. Fix a number $\vartheta$ with $|\vartheta|\le\sqrt b$ and put $x=\rho_n+\eps_n\vartheta\,\xi_{n+1}$. Then $\E[x]=\rho_n$, and
\begin{equation*}
x^-\le\big(\eps_n\sqrt b\,|\xi_{n+1}|-\rho_n\big)_+\le\eps_n\sqrt b\,|\xi_{n+1}|\;\mathbf 1\Big\{|\xi_{n+1}|>\frac{\rho_n}{\eps_n\sqrt b}\Big\}\le\eps_n\sqrt b\,|\xi_{n+1}|^{3}\Big(\frac{\eps_n\sqrt b}{\rho_n}\Big)^{2}.
\end{equation*}
Hence $\E[|x|]\le\rho_n+2\,b^{3/2}M_3\,\eps_n^{3}/\rho_n^{2}$. Since $\rho_n^{2}=n/(n+1)\ge\tfrac12$, we have $\rho_n^{-3}\le2\sqrt2$ and
\begin{equation*}
\E\big[|\rho_n+\eps_n\vartheta\,\xi_{n+1}|\big]\ \le\ \rho_n\,(1+\delta_n),\qquad \delta_n=4\sqrt2\;b^{3/2}M_3\,(n+1)^{-3/2}.
\end{equation*}
Conditioning on $U,U',\xi_{m+1},\dots,\xi_n$ therefore gives $\E[|\Delta_{n+1}|]\le\rho_n(1+\delta_n)\,\E[|\Delta_n|]$. Hence one step multiplies the expected distance by at most $\rho_n(1+\delta_n)$.

Iterating this bound, for $m\le N$,
\begin{equation*}
\E[|\Delta_N|]\le\prod_{n=m}^{N-1}\rho_n\ \prod_{n=m}^{N-1}(1+\delta_n)\ \E[|\Delta_m|]\ \le\ \sqrt{\frac mN}\;\exp\Big(\sum_{n\ge1}\delta_n\Big)\,W_1(\mu,\mu'),
\end{equation*}
because $\prod_{n=m}^{N-1}\sqrt{n/(n+1)}=\sqrt{m/N}$ and $1+\delta\le e^{\delta}$. Since $\sum_{n\ge1}(n+1)^{-3/2}=\sum_{j\ge2}j^{-3/2}$, the exponential equals $\Lambda$. Hence the two normalized sums come together at the rate $\sqrt{m/N}$, which is \eqref{eq:4.5}.
\end{proof}

\subsection{One step changes the limit law very little}\label{sec:4_5}

\begin{lemma}[one-step defect]\label{lem:6}

For every $\eps\in(0,\tfrac12]$,
\begin{equation}
W_1\big(g\,\mathcal K^{\eps},\,g\big)\ \le\ C_L\,\eps^{3},
\label{eq:4.6}
\end{equation}

\begin{equation}
C_L=a_1\,M_3\Big(33+\frac{83}{2}\,b+\frac6b\Big),\qquad a_1=\int\sigma\,g\,\dd y .
\label{eq:4.7}
\end{equation}

\end{lemma}

\begin{proof}
The estimate follows from the cancellation identity \eqref{eq:2.1} and a third-order expansion of smooth test functions. The truncation and derivative bounds needed to control the remainder are given in Appendix~\ref{app:defect}.
\end{proof}

\subsection{Conclusion of the proof}\label{sec:4_6}

Let $N\ge3$ and $\mathcal L_n=\Law(Y_n)$. By \eqref{eq:4.3}, $\mathcal L_N=\mathcal L_3\mathcal K_{3,N}$. By the triangle inequality for $W_1$,
\begin{equation*}
W_1(\mathcal L_N,g)\le W_1\big(\mathcal L_3\mathcal K_{3,N},\,g\mathcal K_{3,N}\big)+W_1\big(g\mathcal K_{3,N},\,g\big).
\end{equation*}
The first term is at most $\Lambda\sqrt{3/N}\,W_1(\mathcal L_3,g)$ by Lemma \ref{lem:5}. For the second, since $g\mathcal K_{n,N}=(g\mathcal K_n)\mathcal K_{n+1,N}$ and $g\mathcal K_{N,N}=g$, the triangle inequality along the chain $g\mathcal K_{3,N},\dots,g\mathcal K_{N,N}$ gives
\begin{equation*}
W_1\big(g\mathcal K_{3,N},g\big)\le\sum_{n=3}^{N-1}W_1\big((g\mathcal K_n)\mathcal K_{n+1,N},\ g\,\mathcal K_{n+1,N}\big)\le\sum_{n=3}^{N-1}\Lambda\sqrt{\frac{n+1}{N}}\;W_1\big(g\mathcal K_n,g\big),
\end{equation*}
by Lemma \ref{lem:5} with $m=n+1$ (for $n=N-1$ the factor is $\Lambda\ge1$). For $n\ge3$ we have $\eps_n=(n+1)^{-1/2}\le\tfrac12$, so Lemma \ref{lem:6} gives $W_1(g\mathcal K_n,g)\le C_L(n+1)^{-3/2}$. Therefore
\begin{equation*}
W_1\big(g\mathcal K_{3,N},g\big)\le\frac{\Lambda\,C_L}{\sqrt N}\sum_{n=3}^{N-1}\frac{1}{n+1}\le\frac{\Lambda\,C_L}{\sqrt N}\int_3^N\frac{\dd x}{x}=\Lambda\,C_L\,\frac{\ln(N/3)}{\sqrt N}.
\end{equation*}
Adding the two terms gives \eqref{eq:1.5}. Both tend to $0$ as $N\to\infty$: $W_1(\mathcal L_3,g)$ is finite because $\mathcal L_3$ and $g$ have finite first moments (Sec.~\ref{sec:4_2} and Lemma \ref{lem:4}(e)). Convergence in distribution follows from (W3). \qed

\begin{remark}
The proof uses the starting value only through $W_1(\mathcal L_3,g)<\infty$. So \eqref{eq:1.5} holds for any $S_1$ with $\E[|S_1|]<\infty$, independent of $\xi_2,\xi_3,\dots$. The first term of \eqref{eq:1.5} shows that the starting law is forgotten at the rate $N^{-1/2}$.
\end{remark}

\section{Proofs of Proposition~\ref{prop:C} and Corollary~\ref{Corlry:D}}\label{sec:5}

\subsection{Proof of Proposition~\ref{prop:C}}\label{sec:5_1}

The derivative of $b\mapsto q=(1+4b)/(1+2b)$ is $2/(1+2b)^{2}>0$. On $(0,\infty)$, $q\to1$ as $b\to0$ and $q\to2$ as $b\to\infty$, so the map is an increasing bijection onto $(1,2)$. On $(-\tfrac12,0)$, $q\to-\infty$ as $b\to-\tfrac12$ and $q\to1$ as $b\to0$, so it is an increasing bijection onto $(-\infty,1)$. In both cases the inverse is $b=(q-1)/(2(2-q))$.

A $q$-Gaussian on the whole line with $1<q<3$ has the density $\propto[1+(q-1)By^{2}]^{-1/(q-1)}$ with $B>0$; it is positive everywhere and decays like $|y|^{-2/(q-1)}$. For $2\le q<3$ the decay exponent $2/(q-1)$ lies in $(1,2]$. If $b<0$, then $g_b$ vanishes outside $(-L,L)$, so it is not positive everywhere. If $b>0$, then by \eqref{eq:1.2} $g_b(y)\,|y|^{2+1/b}\to Z_b^{-1}b^{-\kappa}$ as $|y|\to\infty$, so $g_b$ decays with the exponent $2+1/b>2$. In neither case is $g_b$ a $q$-Gaussian on the line with $2\le q<3$. The Cauchy law is the case $q=2$. \qed

\begin{remark}
For $b>0$ the variance of $g_b$ is finite exactly when $2+1/b>3$, that is, $b<1$, or $q<5/3$. Theorem~\ref{theor:B} does not need this. The variance of the sum behaves accordingly. Taking expectations in \eqref{eq:1.1} gives $\E[S_{n+1}^{2}]=(1+b/n)\,\E[S_n^{2}]+c$, so that $\E[S_n^{2}]/n\to c/(1-b)$ for $b<1$, while $\E[S_n^{2}]/n$ grows like $c\ln n$ for $b=1$ and like a constant times $n^{b-1}$ for $b>1$. The same holds for the unconditional variance $\E[X_{n+1}^{2}]=c+b\,\E[S_n^{2}]/n$ of the summands. Thus for $b\ge1$ every summand has finite variance, but these variances are not bounded in $n$, and the root-mean-square size of $S_n$ grows faster than $\sqrt n$; the quantiles of $S_n$, in contrast, grow like $\sqrt n$, by the convergence in distribution.
\end{remark}

\subsection{Proof of Corollary~\ref{Corlry:D}}\label{sec:5_2}

Throughout, $b\ge1$, $c_b=c\,(1+b)$ (so that $c_N=c_{b_N}$), $\nu=1+1/b$, and $g_b$ is the limit law \eqref{eq:1.2} with $c$ replaced by $c_b$. Let $f_\nu$ be the density of Student's law with $\nu$ degrees of freedom,
\begin{equation*}
f_\nu(t)=\frac{\Gamma\big(\frac{\nu+1}{2}\big)}{\sqrt{\nu\pi}\;\Gamma\big(\frac\nu2\big)}\Big(1+\frac{t^{2}}{\nu}\Big)^{-\frac{\nu+1}{2}},\qquad f_1(t)=\frac{1}{\pi(1+t^{2})},
\end{equation*}
and let $\gamma_c$ be the Cauchy law with density $c^{-1/2}f_1(z/\sqrt c)=\sqrt c/\big(\pi(c+z^{2})\big)$. The proof combines four facts, shown in turn below: the limit law $g_b$ keeps its scale as $b$ grows, it approaches the Cauchy law, a bounded-Lipschitz distance compares laws without a first moment, and the constants of Theorem~\ref{theor:B} grow slowly enough with $b$.

First, the limit law keeps its scale. By Sec.~\ref{sec:1_1}, $g_b$ is the law of $\sqrt{c_b/(1+b)}\;T_\nu=\sqrt c\;T_\nu$, where $T_\nu$ has density $f_\nu$. So $g_b$ has the density $c^{-1/2}f_\nu(z/\sqrt c)$, with the same scale $\sqrt c$ for every $b$.

Second, the limit law approaches the Cauchy law. For every $t$, $f_\nu(t)\to f_1(t)$ as $\nu\to1$, because the Gamma function is continuous and $\Gamma(1)=1$, $\Gamma(\tfrac12)=\sqrt\pi$. Both $f_\nu$ and $f_1$ are probability densities, so Scheff\'e's lemma gives
\begin{equation}
\int_{-\infty}^{\infty}\big|f_\nu(t)-f_1(t)\big|\,\dd t\ \longrightarrow\ 0\qquad(b\to\infty).
\label{eq:scheffe}
\end{equation}
Hence the density of $g_b$ approaches that of the Cauchy law $\gamma_c$ in the integrated absolute difference as $b\to\infty$.

Third, we need a distance for laws without a first moment. The Cauchy law has no first moment, so $W_1$ cannot be used with it. For probability measures $\mu,\mu'$ on $\mathbb R$ let
\begin{equation}
d_{BL}(\mu,\mu')=\sup\Big\{\int h\,\dd\mu-\int h\,\dd\mu'\ :\ |h|\le1,\ \ |h(x)-h(y)|\le|x-y|\ \text{for all }x,y\Big\}.
\label{eq:5.1}
\end{equation}

The following properties are used. Since $d_{BL}$ is a supremum of differences of integrals, it satisfies the triangle inequality. If $\mu,\mu'\in\mathcal P_1$, then $d_{BL}(\mu,\mu')\le W_1(\mu,\mu')$ by \eqref{eq:4.1}, because the supremum in \eqref{eq:5.1} runs over fewer functions. If $\mu$ and $\mu'$ have densities $f_\mu$ and $f_{\mu'}$, then $d_{BL}(\mu,\mu')\le\int|f_\mu-f_{\mu'}|$, because $|h|\le1$. Finally, if $d_{BL}(\mu_N,\mu)\to0$, then $\int h\,\dd\mu_N\to\int h\,\dd\mu$ for every bounded Lipschitz $h$, and $\mu_N\to\mu$ in distribution by the portmanteau theorem, as in (W3).

Fourth, we bound the constants of Theorem~\ref{theor:B} for large $b$. Apply Theorem~\ref{theor:B} with $c$ replaced by $c_b$. The constant $\Lambda$ in \eqref{eq:4.4} does not depend on $c$. For the other quantities in \eqref{eq:1.5}:

\begin{itemize}
\item (a) For $s>\tfrac12$, the Beta integral gives
\begin{equation}
\int_{-\infty}^{\infty}\big(c_b+by^{2}\big)^{-s}\dd y=c_b^{\frac12-s}\sqrt{\frac\pi b}\;\frac{\Gamma\big(s-\frac12\big)}{\Gamma(s)} .
\label{eq:5.2}
\end{equation}
      With $s=\kappa$ this is $Z_b$; with $s=\kappa-\tfrac12$ it is $\int\sigma\,D^{-\kappa}\dd y$. Hence
\begin{equation*}
a_1=\int\sigma g_b\,\dd y=\sqrt{c_b}\;\frac{\Gamma(\kappa-1)\,\Gamma(\kappa)}{\Gamma\big(\kappa-\frac12\big)^{2}} .
\end{equation*}
      Put $x=\kappa-1=1/(2b)\in(0,\tfrac12]$. Then $\Gamma(x)=\Gamma(1+x)/x\le1/x$, because $\Gamma(1+x)\le1$ for $0\le x\le1$ ($\Gamma$ is convex with $\Gamma(1)=\Gamma(2)=1$); $\Gamma(\kappa)=\Gamma(1+x)\le1$; and $\Gamma(\kappa-\tfrac12)=\Gamma(\tfrac12+x)\ge\Gamma(1)=1$, because $\Gamma$ is decreasing on $(0,1.46)$. Therefore $a_1\le2b\,\sqrt{c_b}$.
\item (b) Since $6/b\le6\le6b$ and $33\le33b$, \eqref{eq:4.7} gives $C_L\le a_1M_3\,(39+41.5)\,b\le161\,M_3\,b^{2}\sqrt{c_b}$.
\item (c) $\E[|Y_1|]=\sqrt{c_b}\,\E[|\xi_1|]\le\sqrt{c_b}$. In \eqref{eq:1.4}, $\rho_n,\eps_n\le1$, $\sigma(y)\le\sqrt{c_b}+\sqrt b\,|y|$, $\E[|\xi_{n+1}|]\le1$, and $\xi_{n+1}$ is independent of $Y_n$, so $\E[|Y_{n+1}|]\le(1+\sqrt b)\,\E[|Y_n|]+\sqrt{c_b}$. Hence $\E[|Y_3|]\le(3+3\sqrt b+b)\sqrt{c_b}$. With $\E_{g}[|Y|]\le a_1/\sqrt b\le2\sqrt b\,\sqrt{c_b}$ (Lemma \ref{lem:4}(e)) and the independent coupling,
\begin{equation*}
W_1\big(\Law(Y_3),g_b\big)\le\E[|Y_3|]+\E_{g}[|Y|]\le(3+5\sqrt b+b)\sqrt{c_b}\le9b\,\sqrt{c_b}.
\end{equation*}

\item (d) Insert (a)--(c) into \eqref{eq:1.5} and use $\sqrt{c_b}\le\sqrt{2b}\,\sqrt c$, $9\sqrt3\le16$, $\ln(N/3)\le\ln N$, $b\le b^{2}$ and $M_3\ge1$:
\begin{equation}
\begin{aligned} W_1\Big(\Law\big(S_N/\sqrt N\big),\,g_b\Big)&\le\Lambda\sqrt{2b}\,\sqrt c\;\frac{16\,b+161\,M_3\,b^{2}\ln N}{\sqrt N}\\[4pt] &\le251\,M_3\sqrt c\;b^{5/2}\,\Lambda\;\frac{1+\ln N}{\sqrt N}. \end{aligned}
\label{eq:5.3}
\end{equation}

\end{itemize}
Hence, for large $b$, the bound of Theorem~\ref{theor:B} grows at most like $b^{5/2}\Lambda$.

Finally, let $b=b_N$. For large $N$, $b_N\ge1$ and $b_N^{3/2}\le\theta\ln N$. Since $\sum_{j\ge2}j^{-3/2}=\zeta(\tfrac32)-1$, \eqref{eq:4.4} gives
\begin{equation*}
\Lambda=\exp\Big(4\sqrt2\,\big(\zeta(\tfrac32)-1\big)M_3\,b_N^{3/2}\Big)\le N^{\chi},\qquad \chi=4\sqrt2\,\big(\zeta(\tfrac32)-1\big)M_3\,\theta=\frac{\theta}{2\theta_*}<\frac12 .
\end{equation*}
By \eqref{eq:5.3} and $b_N^{5/2}\le(\theta\ln N)^{5/3}$,
\begin{equation*}
W_1\Big(\Law\big(S^{(N)}_N/\sqrt N\big),\,g_{b_N}\Big)\le251\,M_3\sqrt c\,(\theta\ln N)^{5/3}\,(1+\ln N)\,N^{\chi-\frac12}\ \longrightarrow\ 0 .
\end{equation*}
By the properties of $d_{BL}$ listed above and since $g_{b_N}$ has the density $c^{-1/2}f_{\nu_N}(z/\sqrt c)$,
\begin{equation*}
\begin{aligned} d_{BL}\Big(\Law\big(S^{(N)}_N/\sqrt N\big),\gamma_c\Big)&\le W_1\Big(\Law\big(S^{(N)}_N/\sqrt N\big),\,g_{b_N}\Big)\\[4pt] &\quad+\int\Big|\frac{1}{\sqrt c}f_{\nu_N}\Big(\frac{z}{\sqrt c}\Big)-\frac{1}{\sqrt c}f_1\Big(\frac{z}{\sqrt c}\Big)\Big|\,\dd z, \end{aligned}
\end{equation*}
with $\nu_N=1+1/b_N$. The substitution $z=\sqrt c\,t$ turns the last integral into $\int|f_{\nu_N}-f_1|\,\dd t$, which tends to $0$ by \eqref{eq:scheffe} because $b_N\to\infty$. So $d_{BL}\to0$, and convergence in distribution follows from the last property of $d_{BL}$ listed above. Hence $S^{(N)}_N/\sqrt N$ converges in distribution to the Cauchy law $\gamma_c$. \qed

\begin{remark}[an equivalent form]
Form the summands \eqref{eq:1.1} with $b=b_N$ and $c$ instead of $c_N$, and call their sums $\tilde S_n$. Then $\sqrt{1+b_N}\,\tilde S_n=S^{(N)}_n$ for all $n$, because both start from $\sqrt{c_N}\,\xi_1$ and $(1+b_N)\big(c+b_N\,\tilde S_n^{2}/n\big)=c_N+b_N\,(1+b_N)\,\tilde S_n^{2}/n$. So the corollary can also be read as: with $c$ fixed, $\sqrt{(1+b_N)/N}\;\tilde S_N$ converges in distribution to $\gamma_c$.
\end{remark}

\begin{remark}[the growth condition]
The condition $b_N^{3/2}\le\theta\ln N$ comes from the crude constant $\Lambda$ in \eqref{eq:4.4}, and it lets $b_N$ grow very slowly: with $M_3=1$, $b_N$ is at most about $1.2$ at $N=10^{10}$, that is, $q_N\approx1.7$. The Cauchy law is therefore a limiting point of the family $g_b$ rather than a regime reached at accessible values of $N$. It is not known to be necessary; a sharper contraction estimate than Lemma \ref{lem:5}, with a constant that grows more slowly with $b$ than $\Lambda$, would allow $b_N$ to grow faster. Appendix~\ref{sec:6_5} checks the two ingredients of the proof separately: for fixed large $b$, the law of $S_N/\sqrt N$ is already at the sampling floor around $g_b$ at $N=10^{4}$, and $g_b$ is within about $0.1/b$ of the Cauchy law in Kolmogorov distance.
\end{remark}

\section{Conclusions}\label{sec:concl}

A feedback rule in which the conditional variance of each new summand depends quadratically on the preceding normalized sum produces a family of universal limits. For fixed positive feedback, $b>0$, the limits are the Student distributions, that is, the $q$-Gaussians with $1<q<2$, including those with infinite variance, $5/3\le q<2$. For $-\tfrac12<b<0$ and bounded noise they are the compactly supported $q$-Gaussians with $q<1$; for $-1\le b\le-\tfrac12$ the limits are compactly supported but are not $q$-Gaussians. The Gaussian law, $q=1$, is recovered at $b=0$, and a triangular-array regime with slowly increasing feedback yields the Cauchy law, $q=2$. Every $q<2$ is therefore obtained for a fixed feedback strength, and $q=2$ as a limit. Proposition~\ref{prop:E} in Appendix~\ref{app:propE} shows that the balance identity singles out quadratic feedback: a $q$-Gaussian with $q<2$ satisfies it only for a quadratic conditional variance, and no $q$-Gaussian with $2\le q<3$ satisfies it for any conditional variance. This is a statement about densities that satisfy the identity; for summands $\sigma(S_n/\sqrt n)\,\xi_{n+1}$ with a general variance function, convergence to such a density is not proved here.

The three central limit theorems can be compared directly. The ordinary theorem concerns summands with finite variance, the generalized central limit theorem of L\'evy, Gnedenko and Kolmogorov~\cite{Levy1937,GK}, which we call L\'evy's theorem for short, summands with infinite variance, and the feedback-driven central limit theorem concerns noise with finite variance and quadratic variance feedback. In the ordinary central limit theorem the summands are independent and identically distributed, the normalization is $\sqrt n$, and the limit is the Gaussian, reached at the rate $n^{-1/2}$ of the Berry--Esseen theorem under a finite third moment. In L\'evy's theorem the summands are independent and identically distributed with regularly varying tails of index $\alpha\in(0,2)$, the normalization is $n^{1/\alpha}$ up to a slowly varying factor, and the limits are the stable laws, whose densities decay like $|y|^{-1-\alpha}$ and have infinite variance; the Cauchy law is the symmetric case $\alpha=1$, reached with the normalization $n$. In the feedback-driven central limit theorem the summands are uncorrelated but not independent: each is a fresh noise variable whose amplitude is set by the current normalized sum. The normalization remains $\sqrt n$, and the limits are the laws $g_b$, with $q=(1+4b)/(1+2b)$ fixed by the feedback strength $b$ in the same way as $\alpha$ is fixed by the tail index of the summands in L\'evy's theorem; the limit does not depend on the law of the noise. The limits have compact support for $q<1$, tails $|y|^{-2/(q-1)}$ for $1<q<2$ and infinite variance for $5/3\le q<2$, although the noise itself may be bounded, so that the heavy tails come from the feedback and not from the noise. The Cauchy law, $q=2$, is reached only as a triangular-array limit in which $b_N\to\infty$ slowly, with the normalization $\sqrt{N/(1+b_N)}$ for fixed $c$ (see the Remark on an equivalent form after Corollary~\ref{Corlry:D}), and for $b>0$ the convergence holds with an explicit bound of order $\ln N/\sqrt N$ in $W_1$ under a finite third moment of the noise, a condition of Lyapunov type that plays the role of the third-moment condition of the Berry--Esseen theorem~\cite{Feller}. The limit itself does not depend on this condition, and infinite variance of the limit arises although every summand has finite conditional variance; for $b\ge1$ the unconditional variances of the summands are finite but unbounded in $n$ (see Remark after Proposition~\ref{prop:C}). The case $b=0$ is the ordinary theorem. Finally, in all three theorems the limit is the law left unchanged, at least to leading order in $1/n$, by one more step of the summation followed by rescaling, and only the step differs: an independent increment of fixed variance preserves the Gaussian, an independent heavy-tailed increment preserves the stable laws, and an increment whose variance $D(y)=c+by^{2}$ depends on the current normalized sum $y$ preserves $g_b$. In the last case the normalization contracts the current value while the new increment spreads it, and the identity $(Dg_b)'=-yg_b$ makes these two effects balance at every point of the limiting density.

In plain terms, the three theorems describe three random walkers. The walker of the ordinary theorem takes independent steps of a fixed typical size; after $n$ steps it is typically at a distance of order $\sqrt n$ from its starting point, and its position, measured in units of $\sqrt n$, is Gaussian. The walker of L\'evy's theorem also takes independent steps, but now and then it makes a very long jump; its excursions are dominated by these rare jumps, and its distance grows faster, like $n^{1/\alpha}$. The walker of the feedback-driven central limit theorem takes a fresh random step with mean zero, as the other two do, but chooses the size of the step from its current position, judged against the distance $\sqrt n$ that a walker would normally have reached after $n$ steps. If it lengthens its steps when it finds itself unusually far out ($b>0$), unusual excursions reinforce themselves and its position develops power-law tails, although every single step is moderate compared with the walker's current distance; this happens even when the noise is bounded. If it shortens its steps when it is far out ($b<0$), it can never pass a certain distance, measured in units of $\sqrt n$, and its position has bounded support. If the size of the step does not depend on the position ($b=0$), it is the walker of the ordinary theorem. In each case the position, measured in units of $\sqrt n$, settles into a fixed shape, and for the feedback walker this shape is $g_b$, which is the $q$-Gaussian with $q=(1+4b)/(1+2b)$ when $b>-\tfrac12$, whatever the law of its random steps.

As a result of the feedback-driven central limit theorem, $q$ is fixed by a measurable property of the process, the strength $b$ with which the normalized sum feeds back on the conditional variance of the next summand, through $q=(1+4b)/(1+2b)$, and it can be predicted before the distribution of the sum is examined. This also distinguishes the $q$-Gaussian mechanism from the L\'evy mechanism, although both produce power-law tails. In L\'evy's theorem the tails of the limit are inherited from heavy-tailed summands, a large sum is typically dominated by a single large summand, and the width of the sum grows like $n^{1/\alpha}$. In the feedback-driven central limit theorem the tails are generated by the feedback; with bounded noise each summand changes the sum by at most a bounded amount plus a fraction of order $n^{-1/2}$ of its current size, so that large sums are built up gradually, and the width of the sum grows like $\sqrt n$ even when the limit has infinite variance, where the width is measured by quantiles, for instance by the interquartile range; for $b\ge1$ the root-mean-square width grows faster (see Remark after Proposition~\ref{prop:C}). Among laws with power-law tails, the two mechanisms meet only at the Cauchy law, which L\'evy's theorem reaches with the normalization $n$; the feedback-driven central limit theorem reaches it only as a triangular-array limit in which the feedback strength $b_N$ grows with $N$, with the normalization $\sqrt{N/(1+b_N)}$ for fixed $c$. Which of them is at work in a given system can therefore be told from the growth of the quantile-based width of the sum and from the way the conditional variance of the increments, given the whole past, depends on the current normalized sum, and not only from the shape of the distribution. The two tests play different roles and are used together. The growth of the quantile-based width separates the feedback mechanism from the L\'evy mechanism, but not from other mechanisms that produce $q$-Gaussians with the normalization $\sqrt n$, such as mixtures of Gaussian laws built from exchangeable variables~\cite{HJU2010}. In such mixtures, given the whole past, the conditional variance of the next increment depends on the earlier increments through the sum of their squares, $X_1^{2}+\dots+X_n^{2}$, and not only through the current normalized sum. Given $S_n$ alone, however, it can have exactly the same quadratic form as in the feedback mechanism: for Gaussian mixtures whose inverse variance has a Gamma law, which have Student marginals, it equals $c+b\,S_n^{2}/n$. The second test must therefore condition on the whole past: the feedback mechanism is singled out when the conditional variance of the next increment, given all earlier increments, depends on them only through $S_n/\sqrt n$. When this dependence is measured, its quadratic coefficient gives $b$, and with it a prediction of $q$ that can be checked against the observed distribution of the sum.

\appendix

\section{Proof of the one-step defect estimate}\label{app:defect}

\begin{proof} Let $Y\sim g$ and $\xi\sim\xi_1$ be independent, $\rho=\sqrt{1-\eps^{2}}$ and $Y^{+}=\rho Y+\eps\,\sigma(Y)\,\xi$, so that $Y^{+}\sim g\mathcal K^{\eps}$. By Lemma \ref{lem:3} it is enough to prove
\begin{equation*}
\big|\E[h(Y^{+})]-\E[h(Y)]\big|\le C_L\,\eps^{3}\qquad\text{for every }h\in\mathcal H.
\end{equation*}
Fix $h\in\mathcal H$, and $R$ with $h'=0$ outside $[-R,R]$. All bounds below use only $|h'|\le1$. The proof truncates the noise, interpolates between $Y$ and $Y^{+}$, expresses the derivative of the interpolation through two integrals $\mathcal A_t$ and $\mathcal B_t$, uses the identity \eqref{eq:2.1} to cancel their leading terms, and bounds the remainders.

We first truncate the noise. Let $s_0=\dfrac{1}{4\sqrt b}$ and $\tilde\xi=\xi\,\mathbf 1\{|\xi|\le s_0/\eps\}$, which need not have mean zero. For $A>0$, $|\xi|\mathbf 1\{|\xi|>A\}\le|\xi|^{3}/A^{2}$ and $\xi^{2}\mathbf 1\{|\xi|>A\}\le|\xi|^{3}/A$. With $A=s_0/\eps$, and using $\E[\xi]=0$, $\E[\xi^{2}]=1$:
\begin{equation}
\begin{gathered} \big|\E[\tilde\xi]\big|\le16\,b\,M_3\,\eps^{2},\qquad 0\le1-\E[\tilde\xi^{2}]\le4\sqrt b\,M_3\,\eps,\qquad \E[|\xi-\tilde\xi|]\le16\,b\,M_3\,\eps^{2},\\[4pt] \E[|\tilde\xi|]\le1,\qquad \E[|\tilde\xi|^{3}]\le M_3 . \end{gathered}
\label{eq:4.8}
\end{equation}
Let $\tilde Y^{+}=\rho Y+\eps\,\sigma(Y)\,\tilde\xi$. Since $h$ is 1-Lipschitz and $Y$ is independent of $\xi$,
\begin{equation}
\big|\E[h(Y^{+})]-\E[h(\tilde Y^{+})]\big|\le\eps\;\E[\sigma(Y)]\;\E[|\xi-\tilde\xi|]\le16\,b\,a_1M_3\,\eps^{3}.
\label{eq:4.9}
\end{equation}
Hence truncating the noise changes the expectation by at most $16\,b\,a_1M_3\,\eps^{3}$.

Next we interpolate between $Y$ and $\tilde Y^{+}$. For $t\in[0,\eps]$ let $\rho_t=\sqrt{1-t^{2}}$ and
\begin{equation*}
Y^{(t)}=\rho_t\,Y+t\,\sigma(Y)\,\tilde\xi,\qquad \Phi(t)=\E[h(Y^{(t)})].
\end{equation*}
Then $Y^{(0)}=Y$ and $Y^{(\eps)}=\tilde Y^{+}$, so $\E[h(\tilde Y^{+})]-\E[h(Y)]=\Phi(\eps)-\Phi(0)$. The derivative of $t\mapsto h(Y^{(t)})$ is $h'(Y^{(t)})\big(\sigma(Y)\tilde\xi-\tfrac{t}{\rho_t}Y\big)$, which is bounded in absolute value by the integrable variable $\sigma(Y)|\tilde\xi|+|Y|$ (note $t/\rho_t\le1$ for $t\le\tfrac12$). By dominated convergence,
\begin{equation*}
\Phi'(t)=\E\Big[h'(Y^{(t)})\Big(\sigma(Y)\,\tilde\xi-\frac{t}{\rho_t}\,Y\Big)\Big].
\end{equation*}
To write $\Phi'$ in a convenient form, we introduce the maps $\Psi$. For $t\in[0,\tfrac12]$ and $|s|\le s_0$ define
\begin{equation*}
\Psi(y)=\Psi_{t,s}(y)=\rho_t\,y+s\,\sigma(y),\qquad \tau(y)=\partial_y\Psi(y)=\rho_t+s\,\sigma'(y).
\end{equation*}
Since $\tfrac{\sqrt3}{2}\le\rho_t\le1$ and $|s\sigma'|\le s_0\sqrt b=\tfrac14$ by Lemma \ref{lem:4}(b), and by Lemma \ref{lem:4}(c),
\begin{equation}
\frac12\le\tau\le\frac54,\qquad |\tau'|=|s|\,\sigma''\le\frac{\sqrt b\,c}{4\,\sigma^{3}},\qquad |\tau''|=|s|\,|\sigma'''|\le\frac{3\,b\,c}{4\,\sigma^{4}} .
\label{eq:4.10}
\end{equation}
In particular $\Psi$ is an increasing bijection of $\mathbb R$. Moreover $\sigma(y)\le\sqrt c+\sqrt b\,|y|$, so $s_0\,\sigma(y)\le\tfrac{\sqrt c}{4\sqrt b}+\tfrac14|y|$ and
\begin{equation*}
|\Psi(y)|\ge\rho_t|y|-s_0\,\sigma(y)\ge\Big(\frac{\sqrt3}{2}-\frac14\Big)|y|-\frac{\sqrt c}{4\sqrt b}\ge\frac12|y|-\frac{\sqrt c}{4\sqrt b}.
\end{equation*}
Since $h'$ is continuous and vanishes outside $[-R,R]$, it vanishes wherever $|\Psi|\ge R$. Hence
\begin{equation*}
h'\big(\Psi_{t,s}(y)\big)=0\quad\text{whenever}\quad|y|\ge R_1:=2R+\frac{\sqrt c}{2\sqrt b},\quad\text{for all such }t,s.
\end{equation*}
Define
\begin{equation*}
\mathcal A_t(s)=\int h'\big(\Psi(y)\big)\,\sigma(y)\,g(y)\,\dd y,\qquad \mathcal B_t(s)=\int h'\big(\Psi(y)\big)\,y\,g(y)\,\dd y .
\end{equation*}
Both integrals run over $[-R_1,R_1]$ only, and the integrands are $C^\infty$ in $(s,y)$. So $\mathcal A_t$ and $\mathcal B_t$ are $C^\infty$ in $s$, derivatives may be taken under the integral, and integrations by parts in $y$ have no boundary terms. Since $|t\tilde\xi|\le\eps\cdot s_0/\eps=s_0$, we have $Y^{(t)}=\Psi_{t,\,t\tilde\xi}(Y)$. Conditioning on $\tilde\xi$, which is independent of $Y\sim g$,
\begin{equation}
\Phi'(t)=\E\big[\tilde\xi\,\mathcal A_t(t\tilde\xi)\big]-\frac{t}{\rho_t}\,\E\big[\mathcal B_t(t\tilde\xi)\big].
\label{eq:4.11}
\end{equation}
Hence $\Phi'$ is expressed through the two integrals $\mathcal A_t$ and $\mathcal B_t$.

We now compute the derivatives of $\mathcal A_t$ and $\mathcal B_t$ in $s$. Let $Q=D\,g=\sigma^{2}g$; by Lemma \ref{lem:1}, $Q'=-y\,g$. Because $\partial_s\Psi=\sigma$ and $\partial_y\Psi=\tau$,
\begin{equation}
\partial_s\big[h'(\Psi)\big]=h''(\Psi)\,\sigma=\frac{\sigma}{\tau}\;\partial_y\big[h'(\Psi)\big].
\label{eq:4.12}
\end{equation}
Using \eqref{eq:4.12} and one integration by parts,
\begin{equation}
\mathcal A_t'(s)=\int\partial_y\big[h'(\Psi)\big]\,\frac{\sigma^{2}g}{\tau}\,\dd y=-\int h'(\Psi)\,\Upsilon\,\dd y,\qquad \Upsilon:=\Big(\frac Q\tau\Big)' .
\label{eq:4.13}
\end{equation}
Differentiate \eqref{eq:4.13} once more. The first term uses \eqref{eq:4.12} and an integration by parts again. For the second, $\partial_s\tau=\sigma'$, so $\partial_s\Upsilon=\partial_y\partial_s(Q/\tau)=-\big(Q\sigma'/\tau^{2}\big)'$. Hence
\begin{equation}
\mathcal A_t''(s)=\int h'(\Psi)\,\Big[\Big(\frac{\sigma\,\Upsilon}{\tau}\Big)'+\Big(\frac{Q\,\sigma'}{\tau^{2}}\Big)'\Big]\dd y .
\label{eq:4.14}
\end{equation}
In the same way,
\begin{equation}
\mathcal B_t'(s)=\int\partial_y\big[h'(\Psi)\big]\,\frac{\sigma\,y\,g}{\tau}\,\dd y=-\int h'(\Psi)\,\Big(\frac{\sigma\,y\,g}{\tau}\Big)'\dd y .
\label{eq:4.15}
\end{equation}
The leading terms cancel. At $s=0$ we have $\Psi(y)=\rho_ty$ and $\tau\equiv\rho_t$, so $\Upsilon=Q'/\rho_t=-y\,g/\rho_t$, and \eqref{eq:4.13} gives
\begin{equation}
\mathcal A_t'(0)=\frac{1}{\rho_t}\int h'(\rho_ty)\,y\,g(y)\,\dd y=\frac{\mathcal B_t(0)}{\rho_t} .
\label{eq:4.16}
\end{equation}
This is where the identity \eqref{eq:2.1} enters, and it is the only place where the specific form of $g$ matters.

It remains to bound the higher derivatives. We bound the integrands in \eqref{eq:4.14} and \eqref{eq:4.15} by multiples of $\sigma g$, using $|h'|\le1$, Lemma \ref{lem:4}, \eqref{eq:4.10}, $1/\tau\le2$, and the consequences $c/\sigma\le\sigma$ and $c/\sigma^{2}\le1$ of Lemma \ref{lem:4}(a).
\begin{itemize}
\item $\Upsilon=\dfrac{Q'}{\tau}-\dfrac{Q\,\tau'}{\tau^{2}}$. Here $\big|\tfrac{Q'}{\tau}\big|\le2|y|g\le\tfrac{2}{\sqrt b}\sigma g$ and $\big|\tfrac{Q\tau'}{\tau^{2}}\big|\le4\,\sigma^{2}g\cdot\tfrac{\sqrt b\,c}{4\sigma^{3}}=\sqrt b\,\tfrac c\sigma\,g\le\sqrt b\,\sigma g$. So $|\Upsilon|\le\big(\tfrac{2}{\sqrt b}+\sqrt b\big)\sigma g$.
\item $\Upsilon'=\dfrac{Q''}{\tau}-\dfrac{2Q'\tau'}{\tau^{2}}-\dfrac{Q\,\tau''}{\tau^{2}}+\dfrac{2Q\,\tau'^{2}}{\tau^{3}}$, with $Q''=-(yg)'$. The four terms are bounded by $4\kappa g$ (Lemma \ref{lem:4}(d)); by $8|y|g\cdot\tfrac{\sqrt bc}{4\sigma^{3}}\le2\tfrac{c}{\sigma^{2}}g\le2g$; by $4\sigma^{2}g\cdot\tfrac{3bc}{4\sigma^{4}}\le3b\,g$; and by $16\sigma^{2}g\cdot\tfrac{bc^{2}}{16\sigma^{6}}\le b\,g$. So $|\Upsilon'|\le(4\kappa+2+4b)\,g$.
\item $\Big(\dfrac\sigma\tau\Big)'=\dfrac{\sigma'}{\tau}-\dfrac{\sigma\tau'}{\tau^{2}}$, bounded by $2\sqrt b+4\sigma\cdot\tfrac{\sqrt bc}{4\sigma^{3}}\le3\sqrt b$. Hence
\begin{equation*}
\Big|\Big(\frac{\sigma \Upsilon}{\tau}\Big)'\Big|\le3\sqrt b\,|\Upsilon|+2\sigma|\Upsilon'|\le\big(6+3b+8\kappa+4+8b\big)\sigma g=(10+11b+8\kappa)\,\sigma g .
\end{equation*}

\item $\Big(\dfrac{Q\sigma'}{\tau^{2}}\Big)'=\dfrac{Q'\sigma'}{\tau^{2}}+\dfrac{Q\sigma''}{\tau^{2}}-\dfrac{2Q\sigma'\tau'}{\tau^{3}}$. The three terms are bounded by $4|y|g\sqrt b\le4\sigma g$; by $4\sigma^{2}g\cdot\tfrac{bc}{\sigma^{3}}\le4b\,\sigma g$; and by $16\sigma^{2}g\sqrt b\cdot\tfrac{\sqrt bc}{4\sigma^{3}}\le4b\,\sigma g$. So the whole is at most $(4+8b)\,\sigma g$.
\item $\Big(\dfrac{\sigma yg}{\tau}\Big)'=\dfrac{\sigma'yg+\sigma(yg)'}{\tau}-\dfrac{\sigma yg\,\tau'}{\tau^{2}}$. Here $|\sigma'yg|=\tfrac{by^{2}}{\sigma}g\le\sigma g$ and $|\sigma(yg)'|\le2\kappa\,\sigma g$, so the first fraction is at most $2(1+2\kappa)\,\sigma g$. The second is at most $4\tfrac{\sigma^{2}}{\sqrt b}g\cdot\tfrac{\sqrt bc}{4\sigma^{3}}\le\sigma g$. So the whole is at most $(3+4\kappa)\,\sigma g$.
\end{itemize}

Integrating, and inserting $8\kappa=8+4/b$ and $4\kappa=4+2/b$, for all $t\in[0,\tfrac12]$ and $|s|\le s_0$:
\begin{equation}
\begin{gathered} |\mathcal A_t''(s)|\le C_A:=\Big(22+19b+\frac4b\Big)a_1,\qquad |\mathcal B_t'(s)|\le C_B:=\Big(7+\frac2b\Big)a_1,\\[4pt] |\mathcal A_t(0)|\le a_1,\qquad |\mathcal B_t(0)|\le\frac{a_1}{\sqrt b}. \end{gathered}
\label{eq:4.17}
\end{equation}
The last two bounds follow from $|h'|\le1$ and Lemma \ref{lem:4}(e).

Finally, we assemble the estimate. By Taylor's formula and \eqref{eq:4.17}, for $|s|\le s_0$,
\begin{equation*}
\begin{aligned} \mathcal A_t(s)&=\mathcal A_t(0)+s\,\mathcal A_t'(0)+r_A(s),&\qquad |r_A(s)|&\le\tfrac12s^{2}C_A;\\[4pt] \mathcal B_t(s)&=\mathcal B_t(0)+r_B(s),&\qquad |r_B(s)|&\le|s|\,C_B . \end{aligned}
\end{equation*}
Insert this into \eqref{eq:4.11} with $s=t\tilde\xi$:
\begin{equation*}
\Phi'(t)=\E[\tilde\xi]\,\mathcal A_t(0)+t\,\E[\tilde\xi^{2}]\,\mathcal A_t'(0)-\frac{t}{\rho_t}\mathcal B_t(0)+\E\big[\tilde\xi\,r_A(t\tilde\xi)\big]-\frac{t}{\rho_t}\E\big[r_B(t\tilde\xi)\big].
\end{equation*}
By \eqref{eq:4.16} the second and third terms combine to $-\dfrac{t}{\rho_t}\big(1-\E[\tilde\xi^{2}]\big)\mathcal B_t(0)$. Now use $t\le\eps$, $1/\rho_t\le2$, \eqref{eq:4.8}, \eqref{eq:4.17} and $M_3\ge1$:
\begin{equation*}
\begin{aligned} |\Phi'(t)|&\le16bM_3\eps^{2}a_1+2\eps\cdot4\sqrt b\,M_3\eps\cdot\frac{a_1}{\sqrt b}+\frac12\eps^{2}M_3C_A+2\eps^{2}C_B\\[4pt] &\le\eps^{2}M_3\Big[(16b+8)\,a_1+\frac12C_A+2C_B\Big]. \end{aligned}
\end{equation*}
This bound holds for every $t\in[0,\eps]$, and $\Phi$ is differentiable there, so the mean value theorem gives
\begin{equation*}
|\Phi(\eps)-\Phi(0)|\le\eps^{3}M_3\Big[(16b+8)\,a_1+\frac12C_A+2C_B\Big].
\end{equation*}
Adding \eqref{eq:4.9},
\begin{equation*}
\big|\E[h(Y^{+})]-\E[h(Y)]\big|\le\eps^{3}\,a_1M_3\Big[(32b+8)+\Big(11+\frac{19}{2}b+\frac2b\Big)+\Big(14+\frac4b\Big)\Big]=C_L\,\eps^{3}.
\end{equation*}
Hence the one-step defect is at most $C_L\,\eps^{3}$, which proves Lemma \ref{lem:6}.

\end{proof}

\section{Quadratic feedback characterizes the $q$-Gaussians}\label{app:propE}

Identity \eqref{eq:2.1} can be read in the opposite direction: given a density, it determines the variance function. The following proposition shows that the $q$-Gaussians are exactly the densities that satisfy the identity with a quadratic variance function, that is, with feedback of the form $c+by^{2}$.

\begin{propositionE}[quadratic feedback characterizes the $q$-Gaussians]\label{prop:E}
Let $0<\ell\le\infty$ and $I=(-\ell,\ell)$. Let $D:I\to(0,\infty)$ be continuous and let $g:I\to(0,\infty)$ be a continuous probability density on $I$ such that $Dg$ is continuously differentiable,
\begin{equation}
\big(D\,g\big)'(y)=-\,y\,g(y)\quad(y\in I),\qquad D(y)\,g(y)\to0\quad(y\to\pm \ell).
\label{eq:2.2}
\end{equation}
Then:
\begin{itemize}
\item[(i)] $\int_I|y|\,g(y)\,\dd y<\infty$, $\int_I y\,g(y)\,\dd y=0$, and
\begin{equation}
D(y)=\frac{1}{g(y)}\int_y^{\ell}t\,g(t)\,\dd t\qquad(y\in I).
\label{eq:2.3}
\end{equation}
In particular, $g$ determines $D$.
\item[(ii)] $D(y)=c+b\,y^{2}$ on $I$ for some $c>0$ and $b\in\mathbb R$ if and only if $I=I_b$ and $g=g_b$, the limit law \eqref{eq:1.2} of the main theorem (the Gaussian with variance $c$ if $b=0$).
\item[(iii)] $g$ is a $q$-Gaussian \eqref{eq:1.3} with $q<2$ and $B>0$ if and only if $D(y)=c+b\,y^{2}$ with $c>0$ and $b>-\tfrac12$; then $q=(1+4b)/(1+2b)$ and $B=(1+2b)/(2c)$. No $q$-Gaussian with $2\le q<3$, in particular not the Cauchy law, satisfies \eqref{eq:2.2} for any $D$.
\end{itemize}
\end{propositionE}

\begin{proof}
(i) Write $Q=Dg>0$. For $0<y<\ell$, $Q'(y)=-y\,g(y)<0$, so $Q$ is decreasing on $[0,\ell)$. Integrating $Q'$ over $[y,R]$ and letting $R\to \ell$, where $Q(R)\to0$ by \eqref{eq:2.2}, gives $Q(y)=\int_y^{\ell}t\,g(t)\,\dd t$ for $0\le y<\ell$. In particular $\int_0^{\ell}t\,g\,\dd t=Q(0)<\infty$. In the same way, $Q(y)=\int_{-\ell}^{y}|t|\,g(t)\,\dd t$ for $-\ell<y\le0$, and $\int_{-\ell}^{0}|t|\,g\,\dd t=Q(0)$. Hence $\int_I|y|\,g\,\dd y=2Q(0)<\infty$ and $\int_I y\,g\,\dd y=Q(0)-Q(0)=0$. For $y<0$ this gives $\int_y^{\ell}t\,g\,\dd t=\int_I t\,g\,\dd t-\int_{-\ell}^{y}t\,g\,\dd t=Q(y)$ as well. Dividing $Q$ by $g>0$ gives \eqref{eq:2.3}.

(ii) Suppose $D=c+by^{2}$ on $I$. The function $Q=Dg$ solves the linear equation $Q'=-\dfrac{y}{D}\,Q$, so
\begin{equation*}
Q(y)=Q(0)\exp\Big(-\int_0^{y}\frac{t}{c+bt^{2}}\,\dd t\Big)=
\begin{cases} Q(0)\,\big(D(y)/c\big)^{-1/(2b)}, & b\ne0,\\[2pt] Q(0)\,e^{-y^{2}/(2c)}, & b=0.\end{cases}
\end{equation*}
If $b\ge0$, then $Q(y)\to0$ only as $|y|\to\infty$, so $\ell=\infty$. If $b<0$, then $D>0$ on $I$ requires $\ell\le\sqrt{c/|b|}$, and $Q=Q(0)(D/c)^{1/(2|b|)}$ tends to $0$ at $\pm \ell$ only if $D(\pm \ell)=0$, that is, $\ell=\sqrt{c/|b|}$. In all cases $I=I_b$, and $g=Q/D$ is a constant multiple of $D^{-1-1/(2b)}$ (of $e^{-y^{2}/(2c)}$ if $b=0$), hence $g=g_b$ after normalization. Conversely, $g_b$ satisfies \eqref{eq:2.2} with $D=c+by^{2}$ by Lemma~\ref{lem:1} (for $b=0$ directly), and by (i) no other $D$ does.

(iii) Let $g\propto v^{1/(1-q)}$ with $v(y)=1-(1-q)By^{2}$, $q<2$, $q\ne1$, on the interval where $v>0$. Substituting $s=v(t)$, $\dd s=-2(1-q)Bt\,\dd t$,
\begin{equation*}
\int_y^{\ell}t\,v(t)^{\frac{1}{1-q}}\,\dd t=\frac{v(y)^{\frac{2-q}{1-q}}}{2B(2-q)},
\end{equation*}
because $v^{(2-q)/(1-q)}\to0$ at the end of the interval: $v\to0$ there if $q<1$, and $v\to\infty$ with a negative exponent if $1<q<2$. By \eqref{eq:2.3},
\begin{equation*}
D(y)=\frac{v(y)}{2B(2-q)}=c+by^{2},\qquad c=\frac{1}{2B(2-q)},\qquad b=\frac{q-1}{2(2-q)}>-\frac12 .
\end{equation*}
For $q=1$, $g\propto e^{-By^{2}}$ and \eqref{eq:2.3} gives $D=1/(2B)$, which is the same formula. One checks directly that $g$ satisfies \eqref{eq:2.2} with this $D$. Solving for $q$ and $B$ gives $q=(1+4b)/(1+2b)$ and $B=(1+2b)/(2c)$. Conversely, if $D=c+by^{2}$ with $b>-\tfrac12$, then $g=g_b$ by (ii), and $g_b$ is the $q$-Gaussian \eqref{eq:1.3}. Finally, for $2\le q<3$ the $q$-Gaussian is positive on $\mathbb R$ and decays like $|y|^{-2/(q-1)}$ with $2/(q-1)\le2$, so $\int|y|\,g\,\dd y=\infty$, which (i) excludes.
\end{proof}

Proposition~\ref{prop:E} is a statement about the identity \eqref{eq:2.2}, not about convergence. For summands $X_{n+1}=\sigma(S_n/\sqrt n)\,\xi_{n+1}$ with a general variance function $D=\sigma^{2}$, the first-order computation after Lemma~\ref{lem:1} indicates that a limit density, if it exists, satisfies \eqref{eq:2.2}; a proof of convergence for general $\sigma$ is not given here. Within this class the proposition identifies the mechanism: the $q$-Gaussians are the densities balanced by quadratic feedback, and by no other variance function. It also shows why the Cauchy law requires a feedback strength that grows with the number of terms (Corollary~\ref{Corlry:D}): no fixed variance function balances a density without a first moment.

\section{Numerical checks}\label{app:num}

The simulations provide numerical checks of the limiting laws and of the estimates used in the proofs. A Mathematica notebook that verifies the calculations of the proofs, symbolically or numerically, and reproduces the simulations, tables and figures of the paper is available on GitHub at \href{https://github.com/thomasoikonomou1?tab=repositories}{https://github.com/thomasoikonomou1?tab=repositories}.

\subsection{Simulation setup}\label{sec:6_1}

The recursion \eqref{eq:1.1} was simulated with $c=1$, using $200{,}000$ independent paths up to $N=10^{4}$. The noise laws were symmetric two-point, symmetric three-point, uniform, skewed two-point, Gaussian, and centered exponential distributions, each with mean zero and unit variance. Bounded noise was used in the negative-feedback cases, as required by the main theorem; Gaussian, centered exponential and symmetric two-point noise were used for positive feedback.

At selected values of $N$, the empirical distribution of $S_N/\sqrt N$ was compared with $g_b$ using the Kolmogorov distance. The sampling floor was estimated from independent samples drawn directly from $g_b$ and is approximately $0.002$ for this sample size.

\subsection{Convergence to the limit law}\label{sec:6_2}

\begin{table}[t]
\caption{Kolmogorov distance between the law of $S_N/\sqrt N$ and $g_b$, with $200{,}000$ paths. The limit laws are: arcsine ($b=-1$), uniform ($b=-\tfrac12$), parabola ($b=-\tfrac14$, $q=0$), and Student laws with $4$, $2$ and $\tfrac32$ degrees of freedom ($b=\tfrac13,1,2$; $q=\tfrac75,\tfrac53,\tfrac95$).}
\label{tab:1}
\centering\small\setlength{\tabcolsep}{4pt}
\begin{tabular}{llllllllll}
\toprule
$b$ & noise & $N=10$ & 30 & 100 & 300 & 1000 & 3000 & $10^{4}$ & floor \\
\midrule
$-1$ & $\pm1$ & 0.0106 & 0.0015 & 0.0021 & 0.0024 & 0.0032 & 0.0046 & 0.0029 & 0.0020 \\
$-1/2$ & three-point & 0.0138 & 0.0024 & 0.0023 & 0.0027 & 0.0016 & 0.0013 & 0.0014 & 0.0020 \\
$-1/2$ & $\pm1$ & 0.0344 & 0.0052 & 0.0017 & 0.0032 & 0.0029 & 0.0025 & 0.0016 & 0.0021 \\
$-1/4$ & uniform & 0.0040 & 0.0010 & 0.0025 & 0.0016 & 0.0014 & 0.0022 & 0.0015 & 0.0019 \\
$-1/4$ & skewed two-point & 0.1208 & 0.0452 & 0.0156 & 0.0060 & 0.0032 & 0.0028 & 0.0030 & 0.0018 \\
$1/3$ & Gaussian & 0.0044 & 0.0021 & 0.0012 & 0.0011 & 0.0024 & 0.0012 & 0.0021 & 0.0019 \\
$1/3$ & centered exponential & 0.0407 & 0.0235 & 0.0127 & 0.0076 & 0.0030 & 0.0022 & 0.0024 & 0.0017 \\
1 & Gaussian & 0.0124 & 0.0049 & 0.0021 & 0.0014 & 0.0021 & 0.0038 & 0.0026 & 0.0024 \\
1 & centered exponential & 0.0424 & 0.0172 & 0.0060 & 0.0036 & 0.0027 & 0.0017 & 0.0029 & 0.0019 \\
2 & Gaussian & 0.0308 & 0.0099 & 0.0026 & 0.0018 & 0.0016 & 0.0030 & 0.0021 & 0.0019 \\
2 & $\pm1$ & 0.0206 & 0.0062 & 0.0030 & 0.0021 & 0.0018 & 0.0027 & 0.0022 & 0.0020 \\
\bottomrule
\end{tabular}
\end{table}

\begin{figure}[t]
\centering
\includegraphics[width=0.55\textwidth]{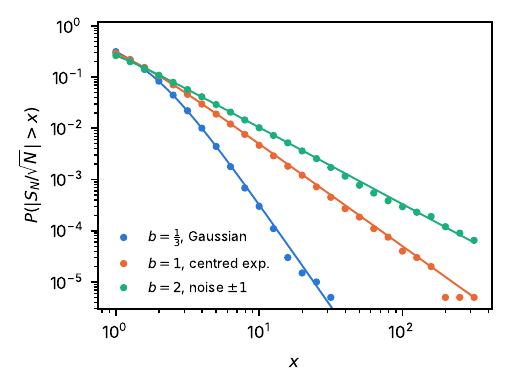}
\caption{Tail probability $P(|S_N/\sqrt N|>x)$ at $N=10^{4}$ on logarithmic axes. The exact tails of $g_b$ decay like $x^{-(1+1/b)}$, that is, with the exponents $4$, $2$ and $\tfrac32$. With $200{,}000$ paths the smallest probability that can be resolved is $5\times10^{-6}$.}
\label{fig:2}
\end{figure}

The Kolmogorov distances are listed in Table~\ref{tab:1}. They decrease overall as $N$ increases, and at $N=10^{4}$ they are within a factor of two of the sampling floor in all eleven cases. Fig.~\ref{fig:2} shows the tails of the law of $S_N/\sqrt N$ for $b>0$, which decay with the exponents of the tails of $g_b$. The convergence is slower for skewed noise, but the final discrepancies are comparable to those obtained from direct samples of $g_b$. The simulations with $b\ge1$ also illustrate why a moment-based argument is insufficient in the heavy-tailed regime: the limiting variance is infinite, although the distribution functions converge.

Panels (c), (e) and (f) of Fig.~\ref{fig:1} come from three additional runs with the same setup: $b=-1/3$ with three-point noise, $b=-1/6$ with uniform noise, and $b=0$ with centered exponential noise. Their Kolmogorov distances to $g_b$ at $N=10^{4}$ are $0.0020$, $0.0015$ and $0.0038$, against sampling floors of $0.0021$, $0.0020$ and $0.0021$. The slightly larger value for $b=0$ is consistent with the slow approach for skewed noise in the ordinary theorem, of order $N^{-1/2}$ by the Berry--Esseen bound; at $N=10$ the distance is $0.041$.

\subsection{Approach to $q=2$}\label{sec:6_5}

Corollary~\ref{Corlry:D} combines two facts: for each fixed $b$, the law of $S_N/\sqrt N$ approaches $g_b$ (Theorem~\ref{theor:B}), and $g_b$ approaches the Cauchy law as $b\to\infty$ (proof of Corollary~\ref{Corlry:D}). The growth condition on $b_N$ is too slow to be followed in a simulation, so the two facts are checked separately. The sums were simulated with $c_b=1+b$ for $b=5,20,50$ ($q=1.91,1.98,1.99$), with $200{,}000$ paths up to $N=10^{4}$. As shown in the proof of Corollary~\ref{Corlry:D}, the limit $g_b$ is then Student's law with $1+1/b$ degrees of freedom and scale $1$, and the Cauchy law has scale $1$ too. Panel (j) of Fig.~\ref{fig:1} comes from a separate run with the same setup, $b=50$ and noise $\pm1$.

\begin{table}[t]
\caption{Kolmogorov distance of the law of $S_N/\sqrt N$ to $g_b$ and to the Cauchy law. The column ``exact'' is the Kolmogorov distance between $g_b$ and the Cauchy law; ``floor'' is the sampling floor of Appendix~\ref{sec:6_1}.}
\label{tab:5}
\centering\small
\begin{tabular}{llllllll}
\toprule
$b$ & noise & to $g_b$: $N=100$ & 1000 & $10^{4}$ & to Cauchy, $N=10^{4}$ & exact & floor \\
\midrule
5 & Gaussian & 0.0049 & 0.0023 & 0.0030 & 0.0182 & 0.0174 & 0.0021 \\
5 & $\pm1$ & 0.0021 & 0.0017 & 0.0013 & 0.0169 & 0.0174 & 0.0018 \\
20 & Gaussian & 0.0202 & 0.0024 & 0.0022 & 0.0051 & 0.0048 & 0.0022 \\
20 & $\pm1$ & 0.0017 & 0.0014 & 0.0014 & 0.0049 & 0.0048 & 0.0023 \\
50 & Gaussian & 0.1062 & 0.0024 & 0.0012 & 0.0023 & 0.0020 & 0.0018 \\
50 & $\pm1$ & 0.0037 & 0.0017 & 0.0025 & 0.0024 & 0.0020 & 0.0018 \\
\bottomrule
\end{tabular}
\end{table}

\begin{figure}[t]
\centering
\includegraphics[width=0.55\textwidth]{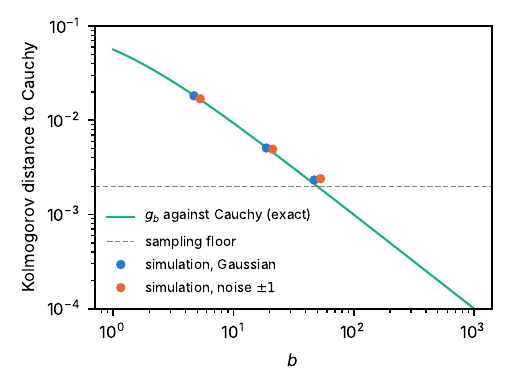}
\caption{Kolmogorov distance between $g_b$ and the Cauchy law as a function of $b$ (line), and the measured distance of the law of $S_N/\sqrt N$ at $N=10^{4}$ to the Cauchy law (points; the two noise laws are drawn slightly apart in $b$). Below $0.002$ a sample of $200{,}000$ paths cannot resolve the difference.}
\label{fig:5}
\end{figure}

The results are listed in Table~\ref{tab:5}. At $N=10^{4}$ the distance to $g_b$ is at the sampling floor in all six runs. The distance to the Cauchy law then equals, within sampling error, the exact distance between $g_b$ and the Cauchy law, which is about $0.1/b$ (Fig.~\ref{fig:5}). For the integrated absolute difference of the densities, used in the proof of Corollary~\ref{Corlry:D}, the factor is about $0.4/b$. The largest value of $|S_N|$ was about $4\times10^{6}$, far from any overflow.

\clearpage
\begin{thebibliography}{99}
\bibitem{Fischer} H.~Fischer, \emph{A History of the Central Limit Theorem: From Classical to Modern Probability Theory} (Springer, New York, 2011).

\bibitem{Lyapunov} A.~M.~Lyapunov, Nouvelle forme du th\'eor\`eme sur la limite de probabilit\'e, M\'em.\ Acad.\ Imp.\ Sci.\ St.-P\'etersbourg \textbf{12}, No.~5 (1901).

\bibitem{Lindeberg} J.~W.~Lindeberg, Math.\ Z.\ \textbf{15}, 211 (1922).

\bibitem{Feller1936} W.~Feller, Math.\ Z.\ \textbf{40}, 521 (1936).

\bibitem{Levy1937} P.~L\'evy, \emph{Th\'eorie de l'addition des variables al\'eatoires} (Gauthier-Villars, Paris, 1937).

\bibitem{GK} B.~V.~Gnedenko and A.~N.~Kolmogorov, \emph{Limit Distributions for Sums of Independent Random Variables} (Addison-Wesley, Cambridge, MA, 1954).

\bibitem{Brown1971} B.~M.~Brown, Ann.\ Math.\ Statist.\ \textbf{42}, 59 (1971).

\bibitem{McLeish1974} D.~L.~McLeish, Ann.\ Probab.\ \textbf{2}, 620 (1974).

\bibitem{Feller} W.~Feller, \emph{An Introduction to Probability Theory and Its Applications}, Vol.~II, 2nd ed. (Wiley, New York, 1971).

\bibitem{Tsallis1988} C.~Tsallis, J.\ Stat.\ Phys.\ \textbf{52}, 479 (1988).

\bibitem{Tsallis} C.~Tsallis, \emph{Introduction to Nonextensive Statistical Mechanics: Approaching a Complex World} (Springer, New York, 2009).

\bibitem{UTS2008} S.~Umarov, C.~Tsallis, and S.~Steinberg, Milan J.\ Math.\ \textbf{76}, 307 (2008).

\bibitem{Bryc2001} W.~Bryc, Commun.\ Math.\ Phys.\ \textbf{219}, 259 (2001).

\bibitem{HJU2010} M.~G.~Hahn, X.~Jiang, and S.~Umarov, J.\ Phys.\ A: Math.\ Theor.\ \textbf{43}, 165208 (2010).

\bibitem{BHW} G.~K.~Basak, I.~Hu, and C.-Z.~Wei, Stochastic Process.\ Appl.\ \textbf{68}, 65 (1997).

\bibitem{Villani} C.~Villani, \emph{Optimal Transport: Old and New} (Springer, Berlin, 2009).

\bibitem{Billingsley} P.~Billingsley, \emph{Convergence of Probability Measures}, 2nd ed. (Wiley, New York, 1999).

\end{thebibliography}
\end{document}